\documentclass[twocolumn]{aastex63}
\usepackage[normalem]{ulem}   
\usepackage{nameref}

\usepackage{amsmath}
\usepackage{amssymb}
\usepackage{mathrsfs}
\usepackage{animate}
\usepackage{graphicx}
\usepackage{rotating}
\usepackage{bbm} 
\usepackage{amsthm}        
\usepackage{comment}

\usepackage{svg}
\usepackage[outline]{contour}
\usepackage{hyperref}
\usepackage{subfigure}
\usepackage{tikz}
\usepackage{mdframed}

\usepackage{lineno}

\newcommand{\p}[1]{{\color{magenta}{#1}}}

\newcommand{\Tp}{T_\mathrm{p}}
\newcommand{\Te}{T_\mathrm{e}}

\newcommand{\FBr}{F_\mathrm{B_R}}
\newcommand{\Fmass}{F_\mathrm{mass}}

\newcommand{\fimf}{f_\mathrm{IMF}}

\newcommand{\finj}{f_\mathrm{inj}}
\newcommand{\Tinj}{\tau_\mathrm{inj}}
\newcommand{\rs}{r_\odot}

\newcommand{\gamp}{\gamma_\mathrm{p}}
\newcommand{\game}{\gamma_\mathrm{e}}

\newcommand{\Ethp}{E_\mathrm{th}^{\mathrm{(p)}}}
\newcommand{\Ethe}{E_\mathrm{th}^{\mathrm{(e)}}}

\newcommand{\Ec}{E_\mathrm{k}}
\newcommand{\Eg}{E_\mathrm{g}}
\newcommand{\Ew}{E_\mathrm{w}}

\newcommand{\Ewii}{E_{\mathrm{w}_2}}

\newcommand{\Eso}{E_{\mathrm{S}_0}}

\newcommand{\Etot}{E_\mathrm{tot}}

\newcommand{\Etotpred}{E_\mathrm{tot|5 au}}

\newcommand{\dv}{\delta v}

\newcommand{\dB}{\delta \mathrm{B}}

\newcommand{\db}{\delta b}

\newcommand{\spani}{\mathrm{SPAN-I}}
\newcommand{\spc}{\mathrm{SPC}}

\newcommand{\std}{\mathrm{std}}

\newcommand{\kurt}{{ \mathrm{kurt} }}

\newcommand{\dUvectii}{\delta \mathrm{\textbf{U}}_{2}}

\newcommand{\dHii}{\delta \mathrm{H}_{2}}

\newcommand{\dBvect}{\delta \mathrm{\textbf{B}}}
\newcommand{\dbvect}{\delta \mathrm{\textbf{b}}}
\newcommand{\dvvect}{\delta \mathrm{\textbf{v}}}

\newcommand{\vvecto}{\mathrm{ \textbf{v}_0 }}
\newcommand{\Bvecto}{\mathrm{ \textbf{B}_0 }}
\newcommand{\bvecto}{\mathrm{ \textbf{b}_0 }}

\newcommand{\vo}{\mathrm{ v_0 }}

\newcommand{\Eps}{\varepsilon}

\newcommand{\Epswii}{\Eps_{\mathrm{w}_2}}

\newcommand{\Evect}{\textbf{E}}

\newcommand{\vvect}{\textbf{v}}
\newcommand{\Bvect}{\textbf{B}}

\newcommand{\Em}{E_\mathrm{m}}

\newcommand{\Wk}{W_\mathrm{k}}
\newcommand{\Wh}{W_\mathrm{H}}
\newcommand{\Ww}{W_\mathrm{W}}
\newcommand{\Wqe}{W_\mathrm{Qe}}
\newcommand{\Wg}{W_\mathrm{G}}
\newcommand{\Wtot}{W_\mathrm{tot}}
\shorttitle{Statistical Energy Budget}
\shortauthors{Dakeyo et al.}

\begin{document}

\title{Statistical Energy Budget of the Solar Wind with Parker Solar Probe and Solar Orbiter~:\\ 1 -- Unbalanced Total Energy Radial Evolution}



\correspondingauthor{Jean-Baptiste Dakeyo}
\email{jbdakeyo@berkeley.edu}

\author[0000-0002-1628-0276]{Jean-Baptiste Dakeyo}
\affiliation{Space Sciences Laboratory, University of California, Berkeley, CA, USA}

\author[0000-0001-8215-6532]{Pascal D\'emoulin}
\affiliation{LIRA, Observatoire de Paris, Universit\'e PSL, CNRS, Sorbonne Universit\'e, Universit\'e de Paris, 5 place Jules Janssen, 92195 Meudon, France}

\author[0000-0002-2916-3837]{Victor Réville}
\affiliation{IRAP, Observatoire Midi-Pyrénées, Universit\'e Toulouse III - Paul Sabatier, CNRS, 9 Avenue du Colonel Roche, 31400 Toulouse, France}

\author[0000-0002-1989-3596]{Stuart Bale}
\affiliation{Space Sciences Laboratory, University of California, Berkeley, CA, USA}
\affiliation{Physics Department, University of California, Berkeley, CA, USA}


\author[0000-0002-8475-8606]{Tamar Ervin}
\affiliation{Space Sciences Laboratory, University of California, Berkeley, CA, USA}
\affiliation{Physics Department, University of California, Berkeley, CA, USA}

\author[0000-0001-6172-5062]{Milan Maksimovic}
\affiliation{LIRA, Observatoire de Paris, Universit\'e PSL, CNRS, Sorbonne Universit\'e, Universit\'e de Paris, 5 place Jules Janssen, 92195 Meudon, France}

\author[0000-0001-6172-5062]{Olga Alexandrova}
\affiliation{LIRA, Observatoire de Paris, Universit\'e PSL, CNRS, Sorbonne Universit\'e, Universit\'e de Paris, 5 place Jules Janssen, 92195 Meudon, France}

\author[0000-0003-2981-0544]{Mingzhe Liu}
\affiliation{Space Sciences Laboratory, University of California, Berkeley, CA, USA}

\author[0000-0002-0396-0547]{Roberto Livi}
\affiliation{Space Sciences Laboratory, University of California, Berkeley, CA, USA}

\author[0000-0002-1128-9685]{Nikos Sioulas}
\affiliation{Imperial College London, South Kensington Campus, London SW7 2AZ, UK}

\author[0000-0002-4559-2199]{Orlando Romeo}
\affiliation{Space Sciences Laboratory, University of California, Berkeley, CA, USA}

\author[0000-0003-4039-5767]{Alexis Rouillard}
\affiliation{IRAP, Observatoire Midi-Pyrénées, Universit\'e Toulouse III - Paul Sabatier, CNRS, 9 Avenue du Colonel Roche, 31400 Toulouse, France}

\begin{abstract}
The magneto-hydrodynamic (MHD) description of the solar wind has long been regarded as one of the most successful frameworks for investigating solar wind heating and acceleration. Under a set of simplifying assumptions, it provides a reduced large-scale description that is widely used to study the energetic role of electromagnetic fluctuations in the solar wind energy budget.
In practice, applying simplification to MHD can result in the loss of some physics compared to an exact set of equations. However, an observational energy budget can help us to verify whether any energetic information has been lost in the process of simplification.
To do so, we directly test the validity of the large-scale non-linear ideal MHD energy budget with solar wind observations from Parker Solar Probe (PSP) and Solar Orbiter (SO).
After verifying the reliability of the measurements through comparison with previous studies of similar solar wind properties, we test the conservation of the total energy predicted by the MHD theory. We find that the average total energy increases with radial distance by $56\%$ ($\pm9\%$) between 14 and 203 solar radii.  
The increase in kinetic energy is not sufficiently compensated by the decrease of the thermal and electromagnetic contributions.
We investigate the major sources of observational uncertainty and find that they cannot account for the observed energy increase. These results suggest that the large-scale theoretical description may neglect an energetically significant contribution to solar wind acceleration. Assuming that such a contribution exists, we predict that this extra energy would scale with radial distance as $r^{-0.54\pm0.38}$.
\end{abstract}


\section{Introduction} \label{sec:Introduction}
A wide range of studies investigating the radial evolution of the solar wind have demonstrated that its dynamics are strongly influenced by ambient electromagnetic (EM) fields \citep{Coleman1968, Alazraki1971, belcher1971, Bale2005, Kasper2019, Bale_2023, Rivera2024, Sioulas2025}.
The role of Alfvénic fluctuations is particularly emphasized, as they are widely regarded as one of the most promising energetic contributors to solar wind heating and acceleration.
Indeed, Alfvénic fluctuations carry energy over a broad range of scales and can affect the solar wind through two distinct mechanisms. At large scales, the radial decrease of the Alfvénic fluctuation energy density generates a wave-pressure gradient that directly accelerates the wind \citep{Alazraki1971, belcher1971}. At smaller scales, wave dissipation transfers energy to particles through kinetic processes, thereby contributing to plasma heating \citep[e.g.][and references therein]{Verscharen2019}. The small scales dissipation processes are expected to directly impact the energy available at large scales, where most of the wave energy is stored.

Nevertheless, simultaneously solving the evolution equations of both particles and large scale EM fluctuations remains computationally demanding and difficult to constrain with observations.
To overcome these limitations, one possible approach is to directly test the MHD equations using observations, provided that all relevant plasma and EM fluctuation properties can be measured \citep{Schwartz1983radial, Halekas2023, Rivera2024, Rivera2025}.

The advent of \textit{Parker Solar Probe} \citep[PSP;][]{fox2016} and \textit{Solar Orbiter} \citep[SO;][]{muller2020} has provided an unprecedented set of in-situ observations, enabling a more detailed investigation of the solar wind's radial evolution. PSP operates closer to the Sun than any previous mission, while SO samples the solar wind at larger heliocentric distances. Together, these missions cover more than a decade in radial distance, making them ideally suited for studying large-scale trends and macroscopic solar wind evolution.

Recent studies have already investigated the radial evolution of the solar wind energy budget using PSP and SO observations, with a particular focus on the energetic role of Alfvénic fluctuations \citep{Halekas2023, Rivera2024, Rivera2025}. For example, \cite{Halekas2023} analyzed the proton radial energy-flux budget using PSP observations between approximately $15~\rs$ and $50~\rs$. They found that, for fast solar wind streams, the relative increase in kinetic energy closely matches the relative decrease in Alfvénic fluctuation energy. In contrast, for slow solar wind streams, the ambipolar electric field, represented by the electron thermal pressure contribution, appears sufficient to explain the observed acceleration.

However, because this analysis was restricted to proton energy budget, and PSP observations between approximately $15~\rs$ and $50~\rs$ (Encounters 4 to 13), both the statistical significance and the radial coverage remain limited. Moreover, the larger-scale radial evolution of the solar wind speed observed between approximately $15~\rs$ and $215~\rs$ using PSP and Helios data \citep{dakeyo2022} places additional constraints on the conclusions of \cite{Halekas2023}. In particular, for slow solar wind streams beyond approximately $100~\rs$, the acceleration associated with the observed proton and electron thermal pressure gradients is no longer sufficient to reproduce the measured increase in solar wind speed \citep{dakeyo2022}. The origin of solar wind acceleration in the inner heliosphere therefore remains only partially understood.

Two recent studies by \cite{Rivera2024} and \cite{Rivera2025} also investigated the energy budget associated with solar wind acceleration. Using Parker spiral alignments between PSP and SO, they examined the radial evolution of protons and electrons in the solar wind between approximately $14~\rs$ and $130~\rs$. Their results are broadly consistent with those of \cite{Halekas2023} regarding the energetic evolution of both fast and slow solar wind streams. However, over these larger radial distances, the total energy flux of the system is not perfectly conserved. 
Since the gravitational energy flux is not computed in a conventional way, one cannot estimate straightforward to what extent the increase is. To obtain a point of comparison, we re-estimated the gravitational flux (and the total flux) from \citep{Rivera2024, Rivera2025} based on the same gravitational contribution used by \cite{Halekas2023} (See Table~\ref{tab:re-estimated_Eflux_Rivera} Appendix~\ref{appendix:re-estime_grav_flux_Rivera}). 
Across the three alignments investigated by \cite{Rivera2024} and \cite{Rivera2025}, a systematic trend is observed in which the total energy measured at SO exceeds that measured at PSP, by 6\%($\pm 20\%$), 55\% ($\pm 18\%$) and 40\% ($\pm35\%$), respectively for the fast, slow and slow Alfvénic wind streams examined in these studies. 
This raises the question of whether these systematic discrepancies originate from observational biases, an incomplete description of the energy budget, or insufficient statistics to establish a significant departure from total energy conservation. A large-scale statistical energy-budget analysis may help clarify these issues.

To address the challenges associated with the role of large-scale Alfvénic fluctuations an the question of total energy conservation during the wind's evolution, we perform a statistical analysis of the large-scale radial energy budget of the solar wind.

In Section~\ref{sec:non_linear_ideal_mhd_equations}, we examine the assumptions underlying the use of Alfvénic fluctuation energy density within the framework of steady-state non-linear ideal MHD. In Section~\ref{sec:SW_properties_radial_evolution}, we assess the reliability of the PSP and SO observations. In Section~\ref{sec:energy_budget}, we confront the theoretical predictions with a large-scale statistical energy budget derived from the observations. Finally, we summarize our results in Section~\ref{sec:conclusion} and discuss their implications in Section~\ref{sec:discussion}.

\section{Solar Wind Properties Radial Evolution}
\label{sec:SW_properties_radial_evolution}

The formalism describing the energy carried by large-scale Alfvén waves is tested against solar-wind observations. We perform an observational analysis of the solar-wind energy budget using measurements from Parker Solar Probe (PSP) and Solar Orbiter (SO).

\subsection{Data collection}

\paragraph{PSP and SO datasets.}
The observational data used in this study are composed of PSP and SO data, taken in between 2020/01/01 to 2025/09/30 and 2020/09/19 to 2025/12/31. 
PSP measurements include partial proton moments from the SPAN-I instrument of the SWEAP suite \citep{Livi2022_SPI_ref, Kasper2015SWEAP}, magnetic field measurements \citep{bale2016, DOI_MAG}, electron density and temperature from quasi-thermal noise (QTN) measurements \citep{QTN_ref_2020}, and electron temperatures from SPAN-E \citep{SPAN_E_ref_2020}. Only measurements obtained below 0.25~au are considered, corresponding to the optimal operational range of SPAN-I.

For SO, we use proton moments from the Proton and Alpha Sensor \citep[PAS,][]{owen_SWA2020} and magnetic field measurements from the MAG instrument \citep{Horbury_MAG2020}. Further details on data selection and processing are provided in \cite{dakeyo2026}.
Electron temperature measurements are not available for SO. We therefore reconstruct $\Te$ using the empirical correlation between $\Te$ and bulk speed $v$ observed in the interplanetary medium \citep[with a Pearson correlation coefficient of -0.6]{dakeyo2026}. This approach provides a realistic estimate of the mean radial trend of $\Te$ within the SO heliocentric distance range. For details on the computation of these synthetic observations, see Appendix~\ref{appendix:subsec_electron_SO_creation}.
Instrumental uncertainties for all measurements are given in Appendix~\ref{appendix:sec_data_preprocessing}. 
Together, the PSP and SO datasets cover a radial range between 9.86~$\rs$ and 218.3~$\rs$.
\subsection{Time Averaging and Fluctuation Computation}
\label{subsec_time_average_fluct_comput}

For the analysis of solar wind properties, plasma parameters are averaged over an adaptive time interval that depends on heliocentric distance $r$, and corresponds to the injection timescales. 
The injection scale is considered to be the length scale at which EM waves are starting to interact non-linearly with the plasma, marking the beginning of the turbulent cascade \citep[so of the inertial range, ][]{Sorriso1999}.

The primary motivation for adopting this scaling is to compare plasma and fluctuation properties at consistent physical scales. In addition, fluctuations measured near the injection scale generally exhibit weak intermittency \citep{Sorriso1999}. Their distributions are therefore expected to be closer to Gaussian, allowing the fluctuation energy to be characterized more robustly by the standard deviation. For a Gaussian distribution, the second-order moment is sufficient to describe the fluctuation energy. In contrast, strongly intermittent distributions require higher-order moments to account for the contribution of the tails.

We adopt the radial scaling $\finj \propto r^{-1}$ \citep{Sioulas2023, Huang2025, dakeyo2026}. The injection frequency is defined as:
\begin{equation}
\finj = 10^{-4} \times \bigg( \frac{r}{r_{1au}} \bigg)^{-1} ,
\end{equation}
where $r_{1au}$ is the Sun--Earth distance in solar radii, capturing the typical $\finj \sim 10^{-4}$~Hz value at 1~au \citep{Alexandrova2013}. The adaptive averaging time is then given by $\Tinj = 1/\finj$. The values of $\Tinj$ range from 6 minutes at PSP’s closest orbit ($\sim 10~\rs$) to 2.5 hours at SO’s farthest orbit ($\sim 215~\rs$).
The $J$-component fluctuation $\delta X_J$ of a given field $\mathbf{X}$ (either $\mathbf{v}$ or $\mathbf{B}$) is computed as the standard deviation of the field over an adaptive time interval of duration $\Tinj$:
\begin{align}
    \langle \delta X_{J}^2 \rangle_{\Tinj} & = \std( X_{J})^2 =  \ \:  \frac{ \sum_{i=1}^{N} \: (X_{J}^{(i)} \: - <X_{J}>_{\Tinj})^2    }{N} ,\label{eq:express_comput_dXj} \\
\text{and} \qquad
    &\langle \delta X \rangle ~=~ \sqrt{ \langle \delta X_R^2 \rangle + \langle\delta X_T^2 \rangle + \langle \delta X_N^2\rangle }, \label{eq:express_comput_dX}
\end{align}
where $N$ is the number of measurements within $\Tinj$, and $J$ denotes each of the $(R,T,N)$ components.
The total fluctuation $\delta X$ is defined as the modulus of the vector fluctuation $\delta \mathbf{X}$.
On board PSP, because the $v_T$ component of the velocity field is subject to more uncertainty than the normal and radial component (private communication with SPAN-Ion team), we approximate the velocity fluctuations to be gyrotropic, i.e. $\langle \dv^2_T \rangle = \langle \dv_N^2 \rangle$ specifically for PSP data. This is supported by the fact that for SO dataset, for which all component are well estimated, $\langle \dv_N \rangle / \langle \dv_T \rangle \approx 1.16$ when computed over all SO data. 

In order to ensure that the statistical energy of the fluctuations follows approximately Gaussian properties, we compute the excess kurtosis, $\kappa$, for each fluctuation distribution evaluated over $\Tinj$. The excess kurtosis provides a measure of the Gaussianity of the distribution, with $\kappa = 0$ expected for a Gaussian distribution. Negative and positive values respectively indicate underrepresented and overrepresented tails compared to a Gaussian distribution. 
Considering a 30\% confidence in the determination of the standard deviation of a distribution, we empirically defined that the range $-1 \leq \kappa \leq 2.7$ corresponds to moderate deviation to Gaussian distributions (See Appendix~\ref{appendix:subsec_inject_scale_filt_kurtosis}). 
Consequently, among the data computed over $\Tinj$, we discard measurements for which $\kappa < -1$ or $\kappa > 2.7$.

\subsection{Description of the observations: Mean radial trends}
\label{subsec:description_obs_mean_trend}

We emphasize that all quantities shown in this section are computed over the injection timescale $\Tinj$, allowing plasma and fluctuation properties to be compared at similar physical scales. For readability, the notation $\langle . \rangle_{\tau}$ is omitted throughout the remainder of this section. \\

\begin{figure*}[t]
\hspace{-0.5cm}
\includegraphics[width=18.5cm]{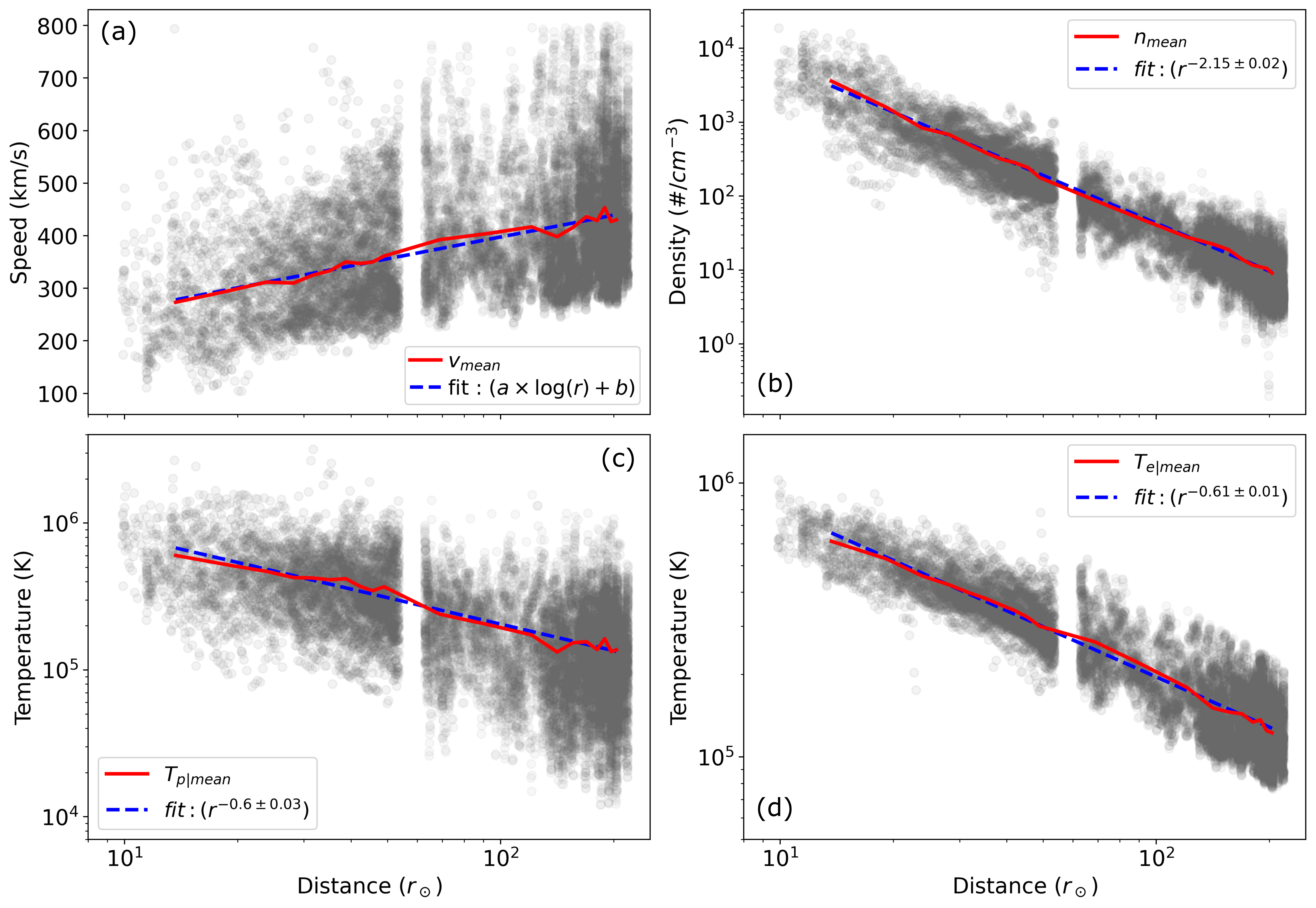}
\caption{Solar wind parameter radial evolution from PSP and SO measurements, with their mean trends. For all panels, each gray dot corresponds to observations averaged over the injection timescale $\Tinj$, and mean trends are represented by solid lines.
Panel~(a): Bulk proton velocity.
Panel~(b): Plasma density, defined as the largest density between $n_p$ and $n_e$, for both PSP and SO datasets. See Appendix~\ref{appendix:sec_data_preprocessing} for more details.
Panel~(c): Proton temperature.
Panel~(d): Electron temperature.
Fits are given in legend for all panels.
For panel (a), the parameters of the fit $a\times log(r) + b$ are~: $a= 60.0 \pm 2.7$, $b= 121.0 \pm 11.4$. For panels (b) to (d), fits corresponds to power law, and the power indexes are given directly in the legend of each panel.
The standard error associated to the mean profiles of all panels (red curves), see Section~\ref{subsec:description_obs_mean_trend}, are~: $\sigma_{v|err} = 1.0\%$, $\sigma_{n|err} = 2.6\%$, $\sigma_{T_{p}|err}= 2.4\%$, and $\sigma_{T_{e}|err} = 0.8\%$.
}
\label{fig_u_Tp_Te_n_all_data_med}
\end{figure*}

Mean trends of the proton bulk velocity, proton, electron, and wave temperatures, plasma density (using QTN electron density for PSP and proton density for SO), and mass flux are computed based on 10 radial bins for PSP and SO, for a total of 20 radial bins covering $\sim 14~\rs$ to $\sim 203~\rs$. Bins are defined independently for PSP and SO data set, and are defined to contain the same number of measurements regarding a given reference data set (PSP or SO observations). This method follows that of \cite{maksimovic2020, dakeyo2022}, but considers a unique mean population rather than median populations.  
Figure~\ref{fig_u_Tp_Te_n_all_data_med} summarizes the radial evolution of the main solar wind parameters.

\paragraph{Standard Errors}
The standard error $\sigma_{X_{bin}|err}$ of a given quantities $X$ when averaged within a given radial bin, is used to define the uncertainty associated to the determination of the mean profiles presented all along the study. It is defined by~:
\begin{align}
    \sigma_{X_{bin}|err} =  \frac{\std (X)}{\sqrt{N}},
\end{align}
where $\std(X)$ is the standard deviation of the value of $X$ within a given radial bin, and $N$ the number of data of this same bin. For convenience of comparing quantities of different orders of magnitude, standard error will be given as a percentage of error. 
Moreover, regarding our data set, since errors in percentage do not vary much with radial distance for whatever the quantity, we average all errors into a unique error estimate for all radial distance, denoted by $\sigma_{X|err}$. 
For a quantity $Y$ composed of different variables $X_i$, the error is computed as~:
\begin{align}
    \sigma_{Y|err} = \sqrt{ \sum_i \bigg(\sigma_{X_i|err} \bigg)^2 } \quad \text{for} \quad Y = \sum_i X_i \:,
\end{align}
\noindent
and
\begin{align}
    \sigma_{Y|err} = |Y| \sqrt{ \sum_i \bigg( \frac{\sigma_{X_i|err}}{X_i} \bigg)^2 } \quad \text{for} \quad Y = \prod_i X_i \:.
\end{align}

\paragraph{Bulk Speed} 
The bulk speed shown in panel~(a) increases with heliocentric distance. An associated log fit of the form $v = a* log(r) + b$ (blue dashed line), shows that the mean wind speed is rising from $278$~km/s at $14~\rs$ to $440$~km/s at $203~\rs$ (from the closest to the farthest radial bin), corresponding to a speed increase of $\sim 162$~km/s. This is consistent with the large-distance acceleration trend observed in previous studies \citep{maksimovic2020, dakeyo2022, halekas2022, dakeyo2026}. When computing the speed increase rate per radial decade $\Delta v = (v(r_{max}) - v(r_{min})) / \log_{10}(r_{max}/r_{min})$ as defined in \cite{dakeyo2026}, this yields a value of $\Delta v \approx 138$~km/s. This result aligns with the estimate of \cite{dakeyo2026} (147~km/s), obtained using a source alignment method, to infer the statistical radial speed increase of individual plasma parcels ($\sim 550$ parcels) in between PSP and SO. Even though the approach of \cite{dakeyo2026} is different to that presented here, it shows that in average, following single parcels of plasma, or taking all observations while mixing streams, provide similar observed speed increase for large statistics and radial coverage. 

\paragraph{Density}
Panel~(b) shows that the plasma density follows approximately $n \propto r^{-2.15\pm0.02}$, slightly steeper than expected for a purely spherical expansion. This could be due to the non-negligible effect of plasma acceleration 
or non-spherical expansion of the interplanetary magnetic flux tube \citep{kopp_holzer1976}. 

\paragraph{Temperatures}
Panel~(c) and (d) shows the mean temperature trends ($\Tp$ and $\Te$), with their respective distribution of observations displayed in the background (gray dots). The proton temperature exhibits an overall radial scaling of $r^{-0.60\pm0.03}$, with two apparent regimes separated by the transition between PSP and SO measurements. Over the PSP radial range, $\Tp$ decreases slightly more slowly than over the SO radial range. This may suggest the presence of two distinct thermal regimes. However, because the transition is not directly observed and coincides exactly with the boundary between the two datasets, any physical interpretation should be treated with caution.

\begin{figure*}[t]
    \hspace{-0.5cm}
    \includegraphics[width=18.5cm]{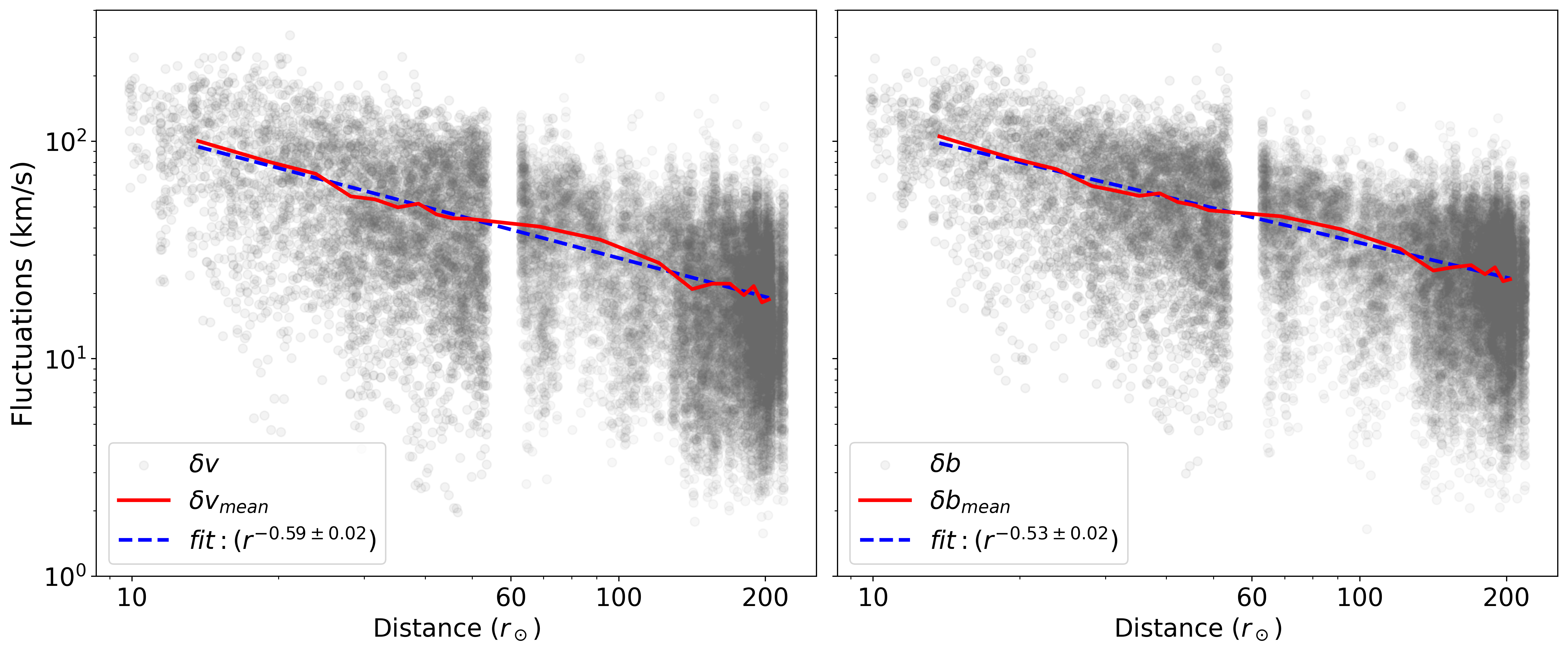}
    \caption{Same as Figure~\ref{fig_u_Tp_Te_n_all_data_med}, for the velocity and magnetic fluctuations computed from Equations~\eqref{eq:express_comput_dXj} and \eqref{eq:express_comput_dX}.
    Left~: Proton bulk speed fluctuation.
    Right~: Alfvén speed fluctuation. 
    The standard error associated to the mean profiles are~: $\sigma_{\dv|err} = 2.4\%$ and $\sigma_{\db|err} = 2.0\%$.
    }
    \label{fig_dv_db_all_data_med}
\end{figure*}

The electron temperature $\Te$ displays a coherent decreasing trend across the PSP and SO distance ranges. Although $\Te$ is reconstructed for SO (see  Appendix~\ref{appendix:subsec_electron_SO_creation}), the overall radial scaling of $r^{-0.61\pm0.01}$ is consistent with known measurements on the same radial interval covered by PSP and Helios measurements \citep{maksimovic2020, dakeyo2022}.
The absence of obvious discontinuities between the PSP measurements and the extrapolated SO values suggests that the reconstructed dataset remains broadly consistent with the expected large-scale evolution of $\Te$.

\paragraph{Polytropic Indexes}
Considering adiabatic indices $\gamp = \game = 5/3$ with a spherical expansion where $n \propto r^{-2.15}$, the expected temperature scalings are $(\Tp, \Te) \propto r^{-1.43}$.
The observed trends, $\Tp \propto r^{-0.60\pm0.02}$ and $\Te \propto r^{-0.61\pm0.01}$, indicate that both species evolve in a sub-adiabatic regime. 

To quantify these polytropic behaviors more precisely, we perform log--log fits between temperature and density, yielding $\gamp = 1.28 \pm 0.01$, $\game = 1.28 \pm 0.01$.
These values illustrate a strong departure from adiabatic cooling ($\gamp = \game = 5/3 \approx 1.66$), supporting the existence of an additional source of particle heating during the solar-wind expansion.   
Although the inferred value of $\game$ remains sensitive to the extrapolated SO temperatures, it agrees well with the range 1.23--1.29 reported in previous studies over similar radial distances \citep{maksimovic2020, dakeyo2022, halekas2022}.
We notice that $\gamp=1.28$ is much lower than $\gamp \approx 1.45$ from \cite{maksimovic2020, dakeyo2022}. 
The major difference is that $\gamp \approx 1.45$ was determined using Helios observations instead of the SO. Accounting for data within the same solar cycle has never been done before. Therefore, even though the obtained value is lower than expected, it may be more temporally coherent. 
The difference in $\gamp$ value could also be due to the fact that the data presented here are a mix of solar minimum and solar maximum measurements. With rising solar activity, the rate of the heating processes can evolve, which may alter the observed polytropic index value when compared to period of minimum activity. Alternatively, the quality of the data on PSP range could also play a role. Even if data have been pre-processed to account for instrumental limitations (See Appendix~\ref{appendix:sec_data_preprocessing}), the field of view of SPAN-I makes it difficult to have the same quality of measurement all along the studied radial range.

\paragraph{Alfvénic Fluctuations}
The measurements of electromagnetic fluctuations are presented in Figure~\ref{fig_dv_db_all_data_med}. Left and right panels show the proton bulk speed fluctuations and the Alfvén speed fluctuations, respectively (as described in Section~\ref{subsec_time_average_fluct_comput}). These exhibit typical values as observed in previous studies with PSP and SO \citep{Ervin2024b, Rivera2024}. Their radial mean pofiles are similar in magnitude, with the average ratio of $(\dv_{mean} / \db_{mean})$ observations being $\sim 0.9$. So they are expected to contain a comparable amount of fluctuating energy.  
The fluctuations exhibit radial scalings of $\dv \propto r^{-0.59 \pm 0.02}$ and $\db \propto r^{-0.53 \pm 0.02}$. The evolution of $\db$ is therefore close to the WKB prediction $\db \propto r^{-0.5}$, suggesting a nearly WKB-like expansion. This is close to what has been previously found in another PSP observation study \citep{Huang2023}.

\begin{figure*}[t]
    \hspace{-0.5cm}
    \includegraphics[width=18.5cm]{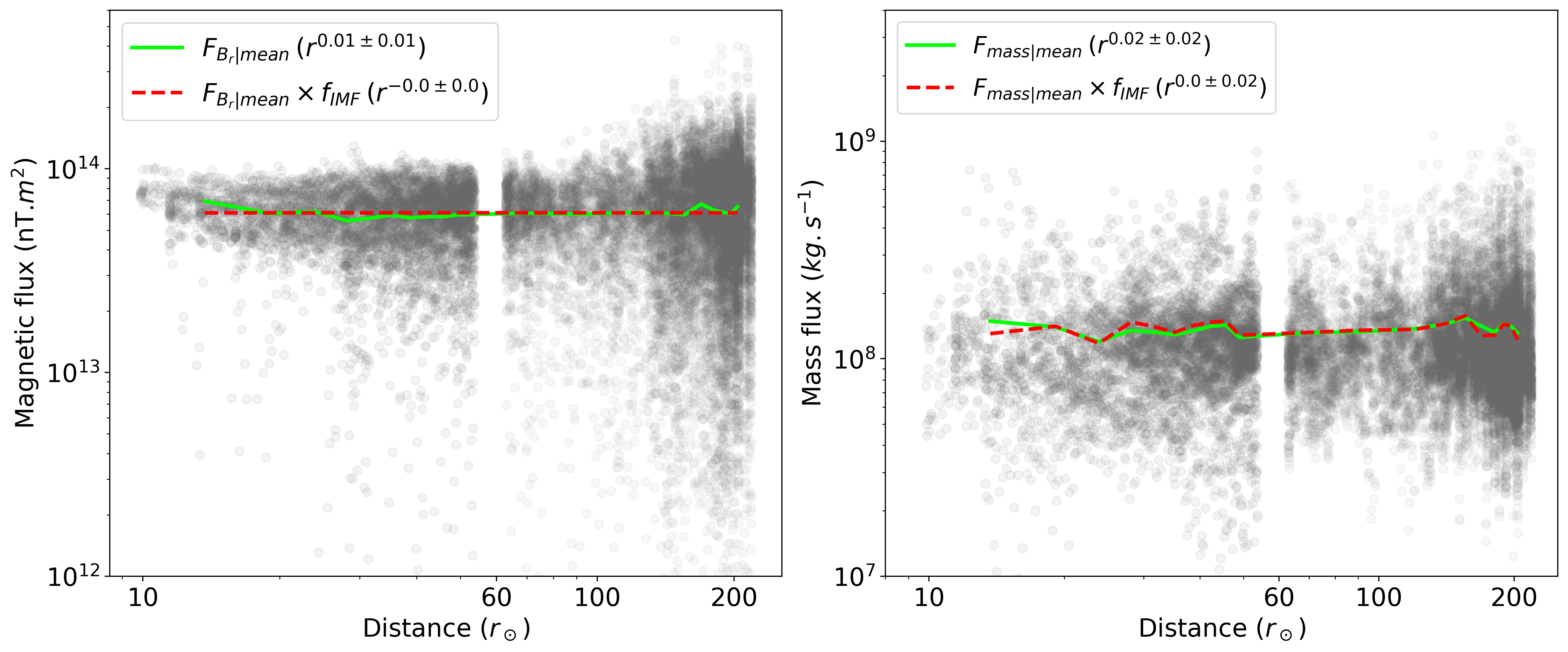}
    \caption{Same as Figure~\ref{fig_u_Tp_Te_n_all_data_med}, for the mass and magnetic fluxes.
    Left: Magnetic flux computed from Equation~\eqref{eq:F_mag}, and 
    Right: Mass flux computed from Equation~\eqref{eq:F_mass} (blue curves). The fluxes corrected by the interplanetary expansion factor $\fimf$, Equation~\eqref{eq:express_fimf}, are shown with red curves.  
    Gray dots corresponds to observations averaged over the injection time scale $\Tinj$.  
    To avoid over-plotting since the power law fits are very close to the mean trend profiles, fitted curves are not displayed, and their values are indicated in legends. 
    The standard error associated to the mean profiles are~: $\sigma_{\Fmass|err} = 2.3\%$ and $\sigma_{\FBr|err} = 1.6\%$. 
    }
    \label{fig_F_mass_mag_all_data_med}
\end{figure*}

\paragraph{Mass and Magnetic Flux}
The magnetic flux $\FBr$ and mass flux $\Fmass$ are defined such as:
\begin{align}
    \FBr =& \: B_R \: r^2, \label{eq:F_mag} \\
    \Fmass =& \ n v_R \: r^2. \label{eq:F_mass}
\end{align}
A non-spherical evolution of $B_R$ (i.e. deviation from $B_R \propto r^{-2}$) is plausibly associated to an additional interplanetary expansion of magnetic flux tube, $\fimf$. The observational estimation of $\fimf$ is~:
\begin{align}
    \fimf = \frac{B_{R|mean}(r=14 \rs) \: \times  (14 \rs)^{2}}{B_{R|mean}(r) \: \times r^{2} } . 
    \label{eq:express_fimf}
\end{align}
This definition is similar to that of the coronal expansion factor \citep{kopp_holzer1976}. Figure~\ref{fig_F_mass_mag_all_data_med} presents the magnetic and mass fluxes respectively on the right and left panels.
The left panel shows that $\FBr$ (green curve) remains nearly constant, with a radial scaling of $r^{0.01 \pm 0.01}$, consistent with the theoretical expectation.
The resulting deviation to $r^{-2}$ defines $\fimf$ (shown in Figure~\ref{fig_fimf} in Appendix~\ref{appendix:complet_fig}). Because the deviation remains small, the resulting values of $\fimf$ remain close to unity.
The correction of $\FBr$ by $\fimf$ (Figure~\ref{fig_F_mass_mag_all_data_med} left panel, red dashed curve), gives a strict conservation of the magnetic flux, according to the definition of $\fimf$. 

The right panel of Figure~\ref{fig_F_mass_mag_all_data_med} shows that $\Fmass$ (blue curve) is also nearly conserved with radial distance and following a radial scaling of $r^{0.02 \pm 0.02}$. The correction of $\Fmass$ by $\fimf$ (red curve) lead actually to a better conservation of corrected mass flux with a radial scaling of $r^{0.00 \pm0.02}$. Additionally, in confirming that the mass flux is well conserved, this allows us to interpret the role played by $v$ in the non-spherical evolution of $n$. 
This implies that the $\sim 162$~km/s increase in wind speed observed in Figure~\ref{fig_u_Tp_Te_n_all_data_med} panel~(a) is responsible for the density decreasing slightly faster than the canonical $r^{-2}$ scaling.

The observed conservation of both magnetic and mass fluxes, which are expected invariants of the flow, strongly supports the overall quality and continuity of the PSP and SO datasets. This gives confidence in the plasma and EM properties used in the large-scale statistical energy-budget analysis presented in the next section.


\section{Non-linear Ideal MHD Equations}
\label{sec:non_linear_ideal_mhd_equations}
In this section, we review the main assumptions commonly adopted to describe the large-scale plasma and EM fluctuation properties of the solar wind, and discuss how they are applied in current studies of the solar-wind energy budget. Ultimately, the most common practice will be observationally tested in Section~\ref{sec:energy_budget} using observations of Section~\ref{sec:SW_properties_radial_evolution}.

The MHD equations describe the main macroscopic properties of a plasma \citep{Bazer1963, belcher1971}. In the case of nonlinear ideal MHD, the plasma evolution is governed by a system of equations relating the mass density $\rho$, the bulk velocity $\vvect$, the thermal pressure $P$, the gravitational potential $\phi$, the magnetic field $\Bvect$, and the electric field $\Evect$~:
\begin{align}
    &\frac{\partial \rho}{\partial t}
    + \nabla . (\rho \vvect) = 0, \label{eq:_MHD_mass_flux} \\
    &\rho\, \frac{\partial \vvect}{\partial t}
    + \rho\, \vvect . \nabla \vvect + \nabla P - \textbf{j} \times \Bvect + \rho \nabla \phi = 0,  \label{eq:momentum_MHD} \\
    &\frac{\partial \Bvect}{\partial t}
    + \nabla \times \Evect = 0,
    \\
    &\Evect = - \vvect \times \Bvect,
    \\
    &\nabla . \Bvect = 0 , \\
    &\nabla \times \Bvect = \mu_0 \ \textbf{j} - \frac{\partial \Evect}{\partial t}, \label{eq:rot_B}
\end{align}
where $\phi = -G M /r$, with $G$ the gravitational constant, $M$ the solar mass, and $r$ the radial distance from the Sun \citep[Equation (4.45) to (4.48)]{Goedbloed2004}. In a single-fluid description, the pressure $P$ represents the combined contribution of all particle species and acts as a single effective pressure driving the plasma expansion.
After manipulating the conservation equations \citep{Bazer1963, Goedbloed2004, Webb2024}, the MHD energy conservation equation is written as~:
\begin{align}
    &\frac{\partial H}{\partial t} + \nabla \cdot \mathbf{U} = 0,
    \label{eq:conserv_energy_flux}
\end{align}
\begin{align}
\text{with} \qquad 
    &H = \frac{1}{2} \rho (\vvect . \vvect) + \frac{P}{\gamma -1} + \rho \phi
    + \frac{\Bvect . \Bvect}{2 \mu_0} , \\
    &\textbf{U} = \bigg( \frac{1}{2} \rho (\vvect . \vvect) +\frac{\gamma \: P}{\gamma -1} + \rho \phi \bigg) \vvect
    + \textbf{S}.
    \label{eq:express_U_H_not_linearized}
\end{align}
where $H$ is the total energy density and $\textbf{U}$ is the total energy flow.

It is worth to emphasize that the electron heat flux influence is not accounted in the present work. While it does contribute to the electron energy flux budget, we face several limitations to include it. Currently, it is not possible to obtain electron heat flux data for SO; these would be more complicated to reconstruct than Te on the SO range; and electron heat flux contribution to the energy budget has been observed to be small compared to the electron enthalpy flux, i.e. the thermal flux \citep{Halekas2023}. Hence, we consider electron heat flux impact to the energy flux budget to be negligible.  
%
The last term of $\textbf{U}$ is the Poynting vector $\textbf{S}$:
\begin{align}
    \textbf{S} = \frac{(\Bvect .\Bvect) \vvect - (\vvect. \Bvect) \Bvect}{\mu_0}
    = - \frac{(\vvect \times \Bvect) \times \Bvect}{\mu_0},
\end{align}
which represents the transport of electromagnetic energy \citep{Goedbloed2004}.


Any field $\mathbf{X}$ can be decomposed as $\mathbf{X} = \textbf{X}_0 + \delta \mathbf{X}$ \citep{Bazer1963}. Here, $\textbf{X}_0 = \langle \mathbf{X} \rangle_\tau$ denotes the background field averaged over the timescale $\tau$, while $\delta \mathbf{X}$ represents the fluctuating component associated with EM waves.

The assumptions of large-scale and incompressible flow can be used to further simplify the description of steady Alfvén-wave evolution. Previous studies 
describe the evolution of Alfvén waves through two conservation equations describing the background flow and the fluctuations \citep[][and references therein]{Bazer1963, Alazraki1971, chandran2009, reville2020, shi2023, Chandran2025}. This approach requires considering wavelengths that remain short compared with the characteristic scale lengths of the background gradients.

In addition, when averaging the fluctuations over an observational timescale $\tau$, the resulting energy flow remains exclusively second order in the fluctuations (denoted by $\delta (.)_{2}$). The energy equation of the fluctuations then becomes \citep{belcher1971}~:
\begin{align}
    &\frac{\partial \langle \dHii \rangle_\tau}{\partial t} + 
    \nabla . \langle \dUvectii \rangle_\tau = 0,\label{eq:express_final_dH2_dU2_flux} 
    \\ \text{with} \quad
    \langle \dHii \rangle_\tau =& \Big\langle \frac{1}{2} \rho \dv^2 + \frac{1}{2} \rho \db^2 \Big\rangle_\tau \: ,\\
    \langle \dUvectii \rangle_\tau =& \Big\langle \frac{1}{2} \rho \dv^2 \: \vvecto
    + \rho  \db^2 \: \vvecto 
    - \rho  (\dvvect. \dbvect) \: \bvecto \Big\rangle_\tau  \: , \label{eq:express_dU2_general}
\end{align}
where $\bvecto = \bvecto / \sqrt{\rho\,\mu_0}$ and $\dbvect = \dBvect / \sqrt{\rho\,\mu_0}$ denote the background magnetic field and magnetic-field fluctuations expressed in units of the Alfvén speed. 

Assuming a WKB expansion with $\dvvect = \mp \dbvect$, these expressions reduce to the forms commonly used in the literature  \citep{Alazraki1971, reville2020, Chandran2025}:
\begin{align}
    \langle \dHii \big\rangle_\tau \approx& \: \rho \langle \db^2 \big\rangle_\tau  = \rho \langle \dv^2 \big\rangle_\tau = \Epswii \:,  \label{eq:express_dH2_wkb} \\
     \langle \dUvectii \rangle_\tau  \approx& \: \Epswii \bigg(\frac{3}{2} \vvecto \pm \bvecto \bigg) \: ,
     \label{eq:express_dU2_wkb}
\end{align}
where $\Epswii$ represents the Alfvén-wave energy density. $\Epswii$ is identified from the expression of $\dHii$ in order to construct a conservation equation analogous to those commonly used for fluid quantities. Indeed, conservation equations require the identification of a quantity that appears consistently in both the temporal and spatial derivative terms and that can be associated with a characteristic transport velocity \citep{Bazer1963, Alazraki1971, belcher1971}. This requires first expressing $\dHii$ explicitly before identifying the corresponding Alfvén-wave energy density within $\dUvectii$, even when considering a stationary expansion of the MHD energy flow.

The role of Alfvén waves in solar-wind acceleration can be inferred from Equations~\eqref{eq:express_final_dH2_dU2_flux}, \eqref{eq:express_dH2_wkb}, and \eqref{eq:express_dU2_wkb} \citep{Alazraki1971}, which can be rewritten as
\begin{align}
     &\nabla . \big[ \  \Epswii   * (\vvecto \pm \bvecto) \ \big] = - \frac{  \Epswii }{2} \ \nabla. \vvecto \quad . \label{eq:energy_flux_alazraki} 
\end{align}
In the absence of dissipation, Equation~\eqref{eq:energy_flux_alazraki} implies that any radial decrease of $\langle \Epswii \rangle_\tau$ on the left-hand side, results in a negative divergence. 
This implies a positive divergence on the right-hand side, $\nabla. \vvecto >0 $, then an increase in the background-flow velocity. Said differently, the background-flow is accelerated by a wave pressure. 
This relation provides an explicit description of the so-called Alfvén-wave pressure gradient \citep{Alazraki1971}.
If wave dissipation processes are included, an additional term equivalent to a heating-energy flux appears on the right-hand side of Equation~\eqref{eq:energy_flux_alazraki} \citep{reville2020, shi2023, Chandran2025}.

This large-scale description of Alfvén-wave evolution, or closely related formulations, has been widely used for several decades to investigate the energetic role of Alfvén waves in the heating and acceleration of the radially expanding solar wind \citep{Alazraki1971,  chandran2009, reville2020, shi2023, Chandran2025}.

\section{Energy Budget Estimation}
\label{sec:energy_budget}

\subsection{Conserved quantity to test}

\begin{figure*}[t]
\hspace{-0.5cm}
\includegraphics[width=18.5cm]{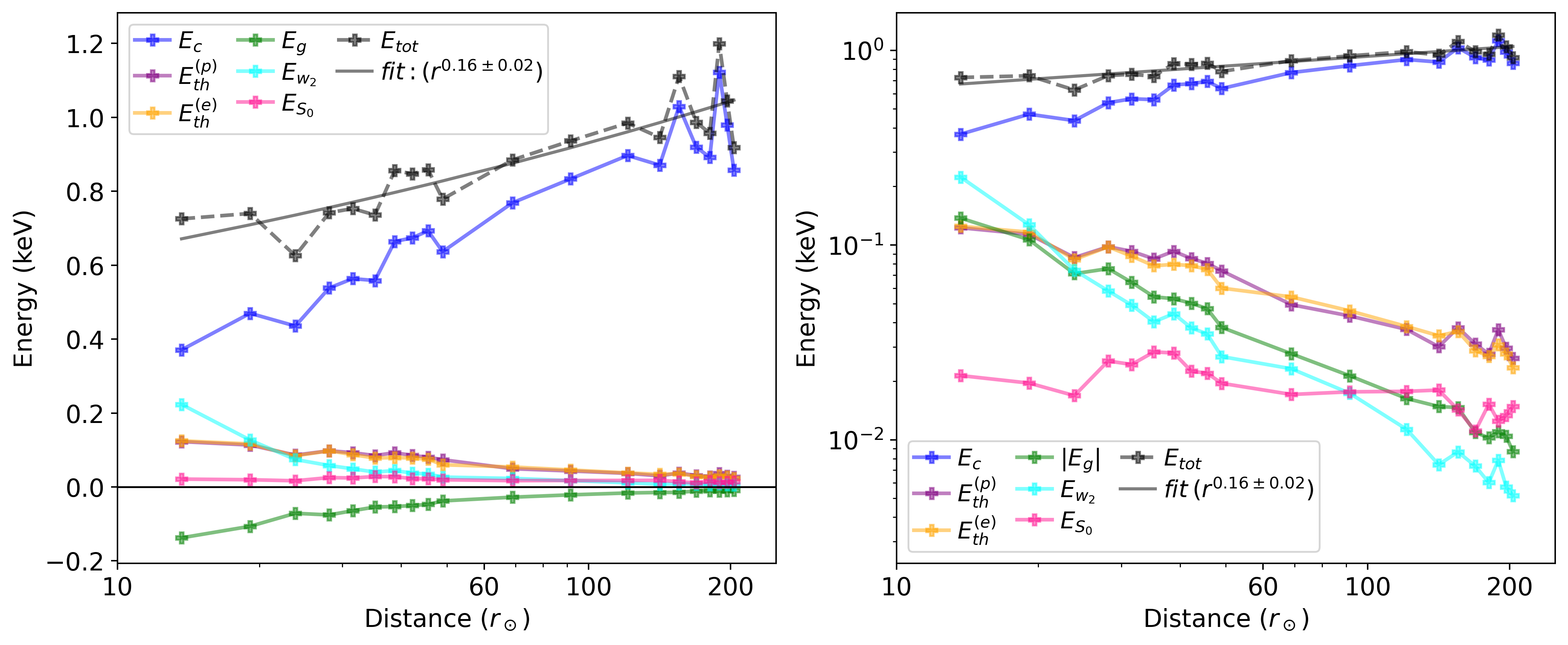}
\caption{Mean trends of the solar wind energy budget contributions, derived from Equations~\eqref{eq:energy_budget_final}, \eqref{eq:U0_final}, \eqref{eq:_dU2_final} and \eqref{eq:express_P_ener_budget}. 
Left panel: Linear scale in ordinates.
Right panel~: Logarithmic scale in ordinates.
The total energy $\Etot$ (black dashed), is the sum of kinetic (blue), proton thermal (violet), 
electron thermal (orange), EM background (pink) and Em fluctuating (cyan) energies. The value of a log-log fit with radial distance of $\Etot$ is displayed by a black solid line, and its corresponding parameters are indicated in the legend panels. 
The standard error associated to the mean profiles are~: $\sigma_{\Ec|err} = 1.4\%$, $\sigma_{\Ethp|err}= 2.4\%$, $\sigma_{\Ethe|err} = 0.8\%$, $\sigma_{\Eg|err} = 0.2\%$, $\sigma_{\Ewii|err} = 2.5\%$, $\sigma_{\Eso|err} = 3.0\%$ and $\sigma_{\Etot|err} = 1.2\%$.
}
\label{fig_energy_all_contrib_lin_log}
\end{figure*}

The energy budget is observationally evaluated based on Equations~\eqref{eq:conserv_energy_flux} to \eqref{eq:express_dU2_general} and the mean radial profiles from PSP and SO observations. 
The energetic contributions are considered for a two-species energy budget (protons and electrons), for which both species have the same speed (i.e. common bulk motion) but different temperatures. Both proton and electron thermal properties are then tested.
To make the energy budget as consistent as possible with the available observations, we assume a stationary evolution of the solar wind on large temporal scales. 
The solar wind is represented as a single mean flux tube, and the energetic properties are studied along this average flow and field line. 

We retain the most general form of the energy equations for fluctuations, Equations~\eqref{eq:express_final_dH2_dU2_flux} and \eqref{eq:express_dU2_general}, for which the condition $\dvvect = \mp \dbvect$ is not imposed. Although Alfvénic fluctuations are expected to contain most of the electromagnetic fluctuating energy, we prefer to separate $\dv$ and $\db$ as we have seen from Figure~\ref{fig_dv_db_all_data_med} that they have slightly different radial scalings and magnitudes.
However, the measurements of velocity fluctuation component $\dv_T$ from PSP embed strong uncertainties because of the SPAN-Ion instrument FOV (see Appendix~\ref{appendix:sec_data_preprocessing}). This makes the direct computation of $\dvvect.\dbvect$ not being reliable on PSP radial range. To overcome this issue, we consider that velocity fluctuations are gyrotropic in good approximation (as in Section~\ref{subsec_time_average_fluct_comput}), and that the scalar product $\dvvect.\dbvect$ can be taken in the $\{R,N\}$ plan (for PSP only) instead of in 3D. This makes the angle $\theta_{\dvvect,\dbvect}$, between $\dvvect$ and $\dbvect$, to be better defined on PSP range and comparable with that on SO range (see Figure~\ref{fig_cos_theta_dvdb} in Appendix~\ref{appendix:complet_fig}). 
Since $\cos(\theta_{\dvvect,\dbvect}) \approx - \text{sign} (b_R) \: |\cos(\theta_{\dvvect,\dbvect})|$ (left panel of Figure~\ref{fig_cos_theta_dvdb}), we consider that $(\dvvect.\dbvect)\bvecto.\textbf{u}_R = - \dv \db \: |\cos(\theta_{\dvvect,\dbvect})| \: |b_R|$ at all time. The mean profile computed for $|\cos(\theta_{\dvvect,\dbvect})|$ is displayed on right panel of Figure~\ref{fig_cos_theta_dvdb} in Appendix~\ref{appendix:complet_fig}. 

The interplanetary expansion factor $\fimf$ is used to correct all fluxes (Section~\ref{subsec:description_obs_mean_trend}), thereby accounting for deviations from a purely spherical expansion of the mean flux tube.
We emphasize that all quantities shown in this section are derived from quantities computed over the injection timescale $\Tinj$. As long as the reliability of plasma properties have been successfully established in Section~\ref{sec:SW_properties_radial_evolution}, they are used to compute any energy quantity presented in this section.
For readability, the notation $\langle . \rangle_{\tau}$ is omitted. \\

To summarize, the conserved quantity of the energy flow equations is expected to be ~:
\begin{align}
     & r^2 \fimf \bigg[  \textbf{U}_{0} . \textbf{u}_R  + \dUvectii. \textbf{u}_R \bigg]  = {\rm constante}, \label{eq:energy_flow_budget_final}
\end{align}
\noindent
with
\begin{align}
    \textbf{U}_0.\textbf{u}_R &= \bigg( \frac{1}{2} \rho v_0^2 + \frac{\gamma}{\gamma-1} P + \rho \phi \bigg) v_R
    + \textbf{S}_{0|R}
    , \label{eq:U0_final} \\
    \dUvectii.\textbf{u}_R &= \frac{1}{2} \rho \dv^2  v_R 
    + \rho  \db^2 v_R
    + \rho   \dv \db \: | \cos(\theta_{\dvvect,\dbvect})| \: |b_R|, \label{eq:_dU2_final} \\
    \textbf{S}_{0|R} &= - \frac{[(\vvecto \times \Bvecto) \times \Bvecto]. \textbf{u}_R}{\mu_0} , \\
    P &= \:  n k_B \Tp + n k_B \Te , \label{eq:express_P_ener_budget}
\end{align}
where $\textbf{u}_R$ is a unit vector oriented along the radial direction, and the total pressure $P$ is the sum of the proton and electron thermal pressure, and $\gamma = 5/3$ in the case of monoatomic particles. 
Since the mass flux, including $\fimf$, is conserved, 
one can rewrite the Equation~\eqref{eq:energy_flow_budget_final} dividing by $\fimf\times \Fmass$, to obtain the energy conservation equation~: 
\begin{align}
 \Etot 
 &= 
  \bigg[  \textbf{U}_{0} . \textbf{u}_R  + \dUvectii. \textbf{u}_R \bigg]  \frac{r^2 }{\Fmass }  \nonumber \\
 &= \Ec + \Ethp + \Ethe + \Eg + \Eso + \Ewii  , \label{eq:energy_budget_final}
\end{align}
\noindent
with
\begin{align}
    \Ec &= \frac{1}{2}\,m\,v_0^2 , \\
    \Ethp &= \frac{5}{2}\, k_B \Tp , \\
    \Ethe &= \frac{5}{2}\, k_B \Te , \\
    \Eg &= -m\,\frac{G M}{r} , \\
    \Eso &= - \textbf{S}_0 . \textbf{u}_R \, \frac{r^2}{\Fmass}  , \\
    \Ewii &= \dUvectii. \textbf{u}_R \, \frac{r^2}{\Fmass} , 
\end{align}
where $m = \rho / n=  m_p+m_e$, is the total mass of a couple proton-electron, of respective mass $m_p$ and $m_e$. $\Etot$ is the mean energy of a couple proton-electron.
Within the framework of the stationary large-scale energy-flow description introduced in Section~\ref{sec:non_linear_ideal_mhd_equations}, the total energy is expected to remain constant with heliocentric distance. Therefore, any statistically significant radial variation of $\Etot$ would indicate either observational biases or limitations of the energetic description itself.\\

\subsection{Energy Budget Observational Results}
 
Figure~\ref{fig_energy_all_contrib_lin_log} shows the radial evolution of all energy contributions, in linear scale for the left panel and logarithmic scale for the right panel. 
The left panel shows that the total energy $\Etot$ increases systematically with heliocentric distance, from approximately 650~eV at the smallest radial distances to about 1000~eV at the largest distances. Fitting the total energy by a log-log fit, it increases proportionally to $r^{0.16 \pm0.02}$, corresponding to a relative increase of 56\% ($\pm 9\%$) between $\sim 14~\rs$ and $\sim 203~\rs$ (from the closest to the farthest mean radial bin). This result suggests that previously reported instances of non-conservation of the total energy flux \citep{Schwartz1983radial, Rivera2024, Rivera2025} may reflect a genuine physical effect rather than isolated observational cases.

The right panel of Figure~\ref{fig_energy_all_contrib_lin_log} further shows that the kinetic energy $\Ec$ is substantially larger than all other energetic contributions throughout the studied radial range. 
The observed increase in kinetic energy is therefore not compensated by decreases in the proton thermal energy $\Ethp$, electron thermal energy $\Ethe$, or fluctuation energy $\Ewii$.

We notice that the electromagnetic energy contained in the Poynting vector (pink curve), remains nearly constant with radial distance. This suggests that the background EM contribution experiences only limited radial evolution, and almost no energy flux exchange. Combined with the near WKB decrease of $\Ewii$, this indicates that in general, EM fields do not dissipate a consequent portion of their energy. 

Comparing our results with those of previous publications summarized in Section~\ref{sec:Introduction} and Appendix~\ref{appendix:re-estime_grav_flux_Rivera}, the observed variation of $\Etot$ correspond in order of magnitude to the increase of the total energy flux (rescaled from Equation~\eqref{eq:_Wtot'_Rivera} to compute similar formalism) found in \cite{Rivera2024, Rivera2025} as follows. 
The fast, slow and slow Alfvénic streams from \cite{Rivera2024, Rivera2025} respectively vary approximately of +6\% ($\pm20\%$ ), +55\% ($\pm18\%$) and +40\% ($\pm35\%$) in total energy, over a radial decade from $\sim$13~$\rs$ to $\sim130$~$\rs$ (Table~\ref{tab:re-estimated_Eflux_Rivera} in Appendix). To make a coherent comparison, we determine the $\Etot$ variation over a radial decade as well, taken in between $\sim$14~$\rs$ and $\sim$140~$\rs$ (black curve in Figure~\ref{fig_energy_all_contrib_lin_log}).  
We obtained a relative variation of $\Etot$ of +48\% over this interval. Considering that our wind speed at 140~$\rs$ is $\sim418$~km/s, then in between the slow (381~km/s) and slow Alfvénic (451~km/s) streams of \cite{Rivera2025}, our +48\% increase is qualitatively consistent with the +55\% and +40\% increases of the slow and slow Alfvénic single stream evolutions.

However, our findings differ from the only previous statistical energy-budget analysis performed using PSP observations \citep{Halekas2023}. For reference, the authors found that the total proton energy budget studied using only PSP data, ranging between $\sim$15~$\rs$ and $\sim$50~$\rs$ (from encounter 4 to 13), was approximately conserved. However, comparing the solar wind trends presented in our study with that of \cite{Halekas2023}, several differences appears. 
Firstly, we consider here a two-species energy budget testing protons and electrons, while \cite{Halekas2023} include the electron energy contributions in the proton energy with the electrical potential, derived from the measured cut-off of electron velocity distributions on the side directed towards the Sun.
Secondly, the qualitative trend of the observed kinetic energy differ in the present study and that of \cite{Halekas2023}.
The average kinetic energy estimated by \cite{Halekas2023} 
shows no significant variation with radial distance on the studied interval (bottom panel of their Figure 1). 
However, the first PSP encounters (E4 to E13) did not sample a sufficiently broad range of solar-wind conditions to provide a representative statistical average. This is visible in the center top panel of Figure~2 from \cite{Halekas2023} which shows the details of their wind population kinetic energy radial evolution. Only slow wind populations appear to be well sampled at all radial distances.  
While all of their wind populations embed non negligible kinetic energy increase (at least $\gtrsim +20\%$ for each), because fast winds are not well sampled at the further distances of PSP, the overall average kinetic energy accounting for all wind population appears to be nearly constant over the entire radial interval \citep[red curve of bottom panel of Figure~1]{Halekas2023}.
This may explain why the average kinetic energy (mixing slow and fast winds) remains unchanged with radial distance, on this time and range of PSP observations.

To verify the difference in the energy conservation with \cite{Halekas2023}, we recomputed $\Etot$ (Equation~\eqref{eq:energy_budget_final}) using $\game=1.28$ instead of $5/3$, with the exact same radial range (below 50~$\rs$, i.e. PSP data only) and restricting to encounters 4 to 13 of PSP (2020/01/29 to 2022/09/30), as done by \cite{Halekas2023}. In such a case, we obtained that $\Etot \propto r^{0.05\pm0.09}$, with a relative variation of +5~\% $\pm 10\%$  between 16~$\rs$ and 46~$\rs$. Given the margin of error, $\Etot$ is actually nearly conserved when restricting to the same data and similar assumptions as in \cite{Halekas2023}. 
Next, we performed the two-species budget with the same dataset as \cite{Halekas2023} but considering $\game = 5/3$ for the electron thermal contribution, as for the protons. 
We obtain $\Etot \propto r^{0.12\pm0.09}$, which is in between \cite{Halekas2023} and present (Figure~\ref{fig_energy_all_contrib_lin_log}) results, while closer to the latter.  
Therefore, our approach is consistent with existing methods and the difference in the results is mainly due to the statistical data samples used, and the $\gamma$ value taken for electron thermal contribution.

\subsection{Source of errors in the Energy Budget}
Investigating observational source of errors, we can mainly identify the use of synthetic electron data from SO, the lack of alpha particles in the energy budget, and possible underestimation of the fluctuations amplitude.

We would like to emphasize that the electron temperature in the SO radial range has been reconstructed from the $v - \Te$ relationship of PSP observations. 
As mentioned in Section~\ref{subsec:description_obs_mean_trend}, the associated polytropic index $\game = 1.28$ agrees well with the range 1.23--1.29 reported in previous studies over similar radial distances \citep{maksimovic2020, dakeyo2022, halekas2022}.
Therefore, one would not expect a drastic change neither of $\game$ nor of the mean profile of $\Te$ if replacing reconstructed data by real in situ ones. 
Moreover, examining the fraction of $\Ethe$ in $\Etot$ on the SO range (Figure \ref{fig_energy_all_contrib_lin_log}), shows that $\Ethe \sim 0.06 \times \Etot$ at most. Even an error of $\pm100\%$ in the reconstructed $\Te$ data can affect $\Etot$ by no more than 6\%, and is therefore far from being able to solve  the observed increase of $\Etot$ by $\pm$56\% ($\pm9$\%). 

Regarding other type of electron observations, the electron heat flux contribution is not accounted for in the present work. This results from observational limitations, but also by the fact that its contribution is expected to be negligible compared to that of $\Ethe$. In order to estimate what impact it would have on $\Etot$, we consider the relative magnitude of the electron heat flux, $q_E$, compared to that of the electron enthalpy flux (equivalent to the ambipolar electric field contribution, $ \phi_E$), from top panel of Figure~1 of \cite{Halekas2023}. On PSP radial range, considering an average wind trend, \cite{Halekas2023} show that, for $\phi_E/q_E \sim 3 - 10$, confirming that electron heat flux is a secondary contribution of the electron energy budget. Including $q_E$ to the energy budget would be equivalent to increase $\Ethe$ by between $10\%$ and $30\%$. For convenience, we retain a value of 20\%. Such a relative increase of $\Ethe$ added to our energy budget results in a change of $\Etot$ variation of $\Etot \propto r^{0.15\pm0.02}$ instead of $\Etot \propto r^{0.16\pm0.02}$, giving a 52\% ($\pm9\%$) variation compared to the 56\% ($\pm9\%$) initially observed. The total energy variation is reduced, but is not qualitatively different, remaining still important. We can conclude that neglecting the electron heat flux indeed has no qualitative impact on the present energy budget results, and that electron heat flux is not in capacity for conciliating the missing energy to $\Etot$.

Next, to evaluate the impact of the lack of alpha particles in the energy budget, we use generally known properties of alphas to understand 
their impact on the kinetic, thermal, and gravitational energy terms. All the details are presented in Appendix~\ref{appendix:sec:alpha_influence_energy_budget}. 
Accounting for typical alpha properties, we obtain $\Etot \propto r^{0.17\pm0.02}$, which is a slightly steeper variation than without accounting alphas ($\Etot \propto r^{0.16 \pm0.02}$), with a relative increase of $\Etot$ of 60\% ($\pm9$\%) instead of the 56\% ($\pm9$\%) originally obtained. 
Including alpha particles therefore strengthens the observed non-conservation rather than alleviating it. 
This is due to the fact that alphas add to the system more kinetic and gravitational energy, than thermal energy. The overall results is the same as for protons, the observed alpha thermal pressure is not large enough to overcome the gravitational attraction, and accelerate the particles as observed. 
We observe that the influence of the alpha particles and electron heat flux cancels out, as their respective impact on the energy budget is approximately opposite and of the same order of magnitude.

Finally, another possible source of uncertainty lies in estimating the fluctuations amplitude. This issue was actually anticipated in Section~\ref{subsec_time_average_fluct_comput} for a reliable energy budget computation. We would like to reiterate that, for any ($\dv, \db$) averaged over the injection time scales, the excess kurtosis of the distribution was computed and only those close enough to Gaussian properties were retained. Therefore, $\Ewii$ already represents the most energetic fluctuations and is less likely to have been systematically underestimated.

\subsection{Estimating the "Missing" Energy Contribution}

After accounting for the dominant identifiable sources of uncertainty, the statistical energy budget remains significantly unbalanced over the full radial interval. 
This leaves open the possibility that the energetic description itself may be incomplete and could neglect one or more energetically significant contributions. 
The effect of this/these missing terms, is of major importance for the solar wind acceleration in the inner heliosphere.

\begin{figure}[t]
\hspace{-0.5cm}
\includegraphics[width=8.5cm]{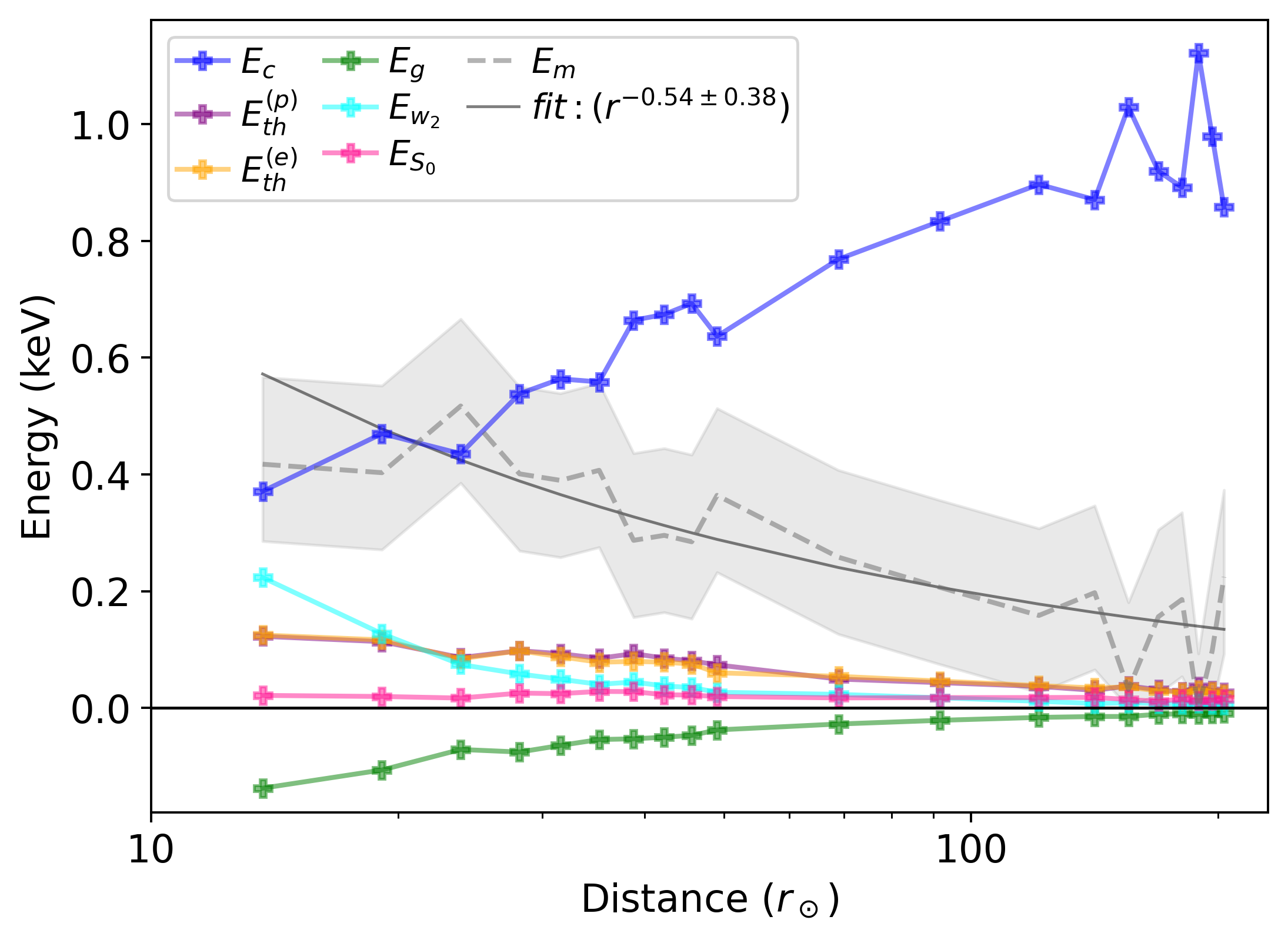}
\caption{Same as left panel of Figure~\ref{fig_energy_all_contrib_lin_log} with the additional predicted "missing" energy contribution $\Em$ (dashed gray), its corresponding uncertainty (shaded gray area), and power law fit (solid black curve). Uncertainties of the fit are estimated by the average deviation of the power law fits coefficients, of the top and bottom lines delimiting the gray shaded area. 
}
\label{fig_energy_flux_r2_Umiss}
\end{figure}

Assuming that an additional contribution $\Em$ restores exact energy conservation, we can estimate the radial evolution that such a term would need to exhibit. To do so, one need to estimate the actual total energy of the system. This can be done by considering distances where wind has completed the main part of its acceleration. Then, the total energy equals to the kinetic energy. The trend observed on panel (a) of Figure~\ref{fig_u_Tp_Te_n_all_data_med} suggests that the bulk speed may keep increasing further 1~au.
Observations of Ulysses mission spanning between 1~au and 5~au, could monitor the last phase of acceleration of the wind. 
For Ulysses observations taken within a similar solar cycle period (i.e. rising maximum), we obtain a speed increase from 429~km/s to 468~km/s, giving a relative increase of $\sim$9\%. Please refer to Appendix~\ref{appendix:average_asympt_wind_speed_pred} for more details. At 1~au the mean bulk speed measured by SO is 440~km/s which is only 11~km/s, or $3\%$, higher than Ulysses estimate while the instruments and the solar cycle are different.   
We know that the wind at least reached a terminal speed of $\geq$ 440~km/s, we expect that it may reach 468~km/s, but it could be larger than that. The present solar cycle and that of Ulysses may embed unaccounted difference. To account for such unknowns, the error on the wind terminal speed is taken as 100\% of the remaining speed increase 
i.e. $\pm$28~km/s. The missing energy is then estimated by~:
\begin{align}
    \Em(r) = \Etotpred - \Etot(r),
\end{align}
with $\Etotpred = \Ec(\vo = 468\pm 28 \ \text{km/s})$. 
The mean profiles of $\Em$ and its associated uncertainty are presented in Figure~\ref{fig_energy_flux_r2_Umiss} by a dashed gray curve, and the gray shaded area, respectively. 
In average $\Em \sim 2.0\, (\pm1.2) \times [\Ethp+\Ethe + \Ewii]$,
which means that the major energetic contribution to the wind acceleration is absent in the current description. 
Regarding the radial scaling, we predict that $\Em$ should approximately follows $\Em \propto r^{-0.54 \pm 0.38}$.

\begin{figure*}[t]
    \hspace{-0.5cm}
    \includegraphics[width=18.5cm]{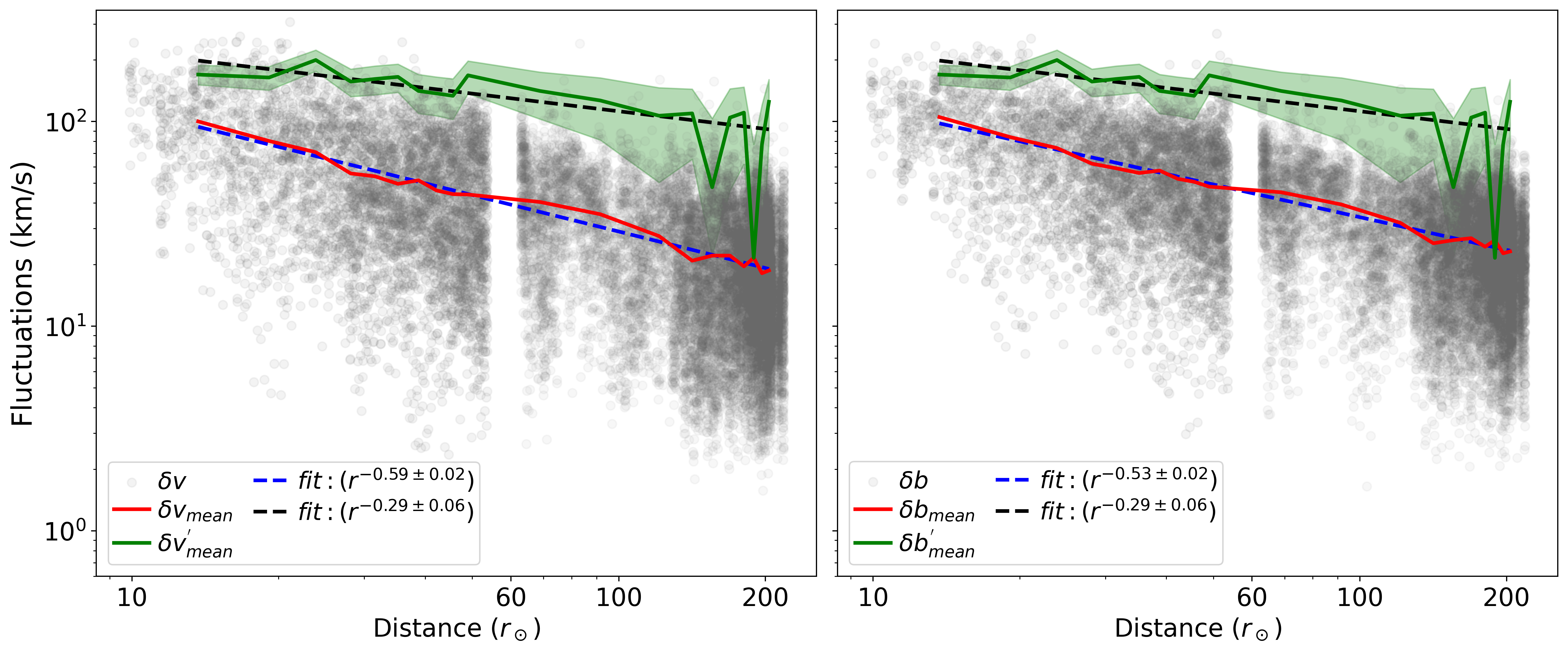}
    \caption{Same as Figure~\ref{fig_dv_db_all_data_med}, but for $\dv$ and $\db$ adjusted to represent the amplitude they would have if $\Ewii$ was able to allow to conserve $\Etot$, considering $\Ewii' = \Em + \Ewii$, and $\dv' = \dv \sqrt{\Ewii'/ \Ewii}$ and $\db' = \db \sqrt{\Ewii'/ \Ewii}$. 
    Left~: Proton bulk speed fluctuation.
    Right~: Alfvén speed fluctuation.
    The standard error associated to $\dv$ and $\db$ are the same as in Figure~\ref{fig_dv_db_all_data_med}. The uncertainty (green shaded area) is determined from that of $\Em$ (gray shaded are in Figure~\ref{fig_energy_flux_r2_Umiss}).
    }
    \label{fig_dv_db_all_data_med_ajust_Emiss}
\end{figure*}

In the hypothetic case where EM fluctuations as described by $\Ewii$, would have been large enough to balance the energy budget, we have estimated a hypothetic EM wave contribution $\Ewii' = \Ewii + \Em$. The resulting equivalent velocity and, respectively $\dv'$ and $\db'$, can be estimated by $\dv' = \dv \sqrt{\Ewii' / \Ewii}$ and $\db' = \db \sqrt{\Ewii' / \Ewii}$ (Figure~\ref{fig_dv_db_all_data_med_ajust_Emiss}). We observe that both the amplitude and the scaling of $\dv'$ and $\db'$ are quite different from $\dv$ and $\db$ from the data. Readjusted curves (green) are approximately twice larger at closest approach, and until 5 times larger a furthest approach than observe trends (red). 
The readjusted scaling is approximately twice as flat with $(\dv', \db') \propto r^{-0.29 \pm 0.06}$, representing a much slower decrease than the dissipation-free WKB scaling. 
Observing such fluctuation scaling, would be possible if EM fluctuations are continuously generated in situ. 
To estimate the required level of energy to be injected, we take the same reference fluctuation amplitude at the closest approach, noted $\db_{14\rs}$. The relative radial dependence is~:
\begin{align}
    \db_{(203\rs)} = \db_{14\rs} \bigg(\frac{203}{14}\bigg)^{-0.53} = 0.24 \times \db_{14\rs} ,  \\
    \db_{(203\rs)}' = \db_{14\rs} \bigg(\frac{203}{14}\bigg)^{-0.29} = 0.46 \times \db_{14\rs} ,
\end{align}
We obtain that $\db_{203\rs}' / \db_{203\rs} \approx 1.9$. Starting from the same reference, a relative factor of $\sim$2 in fluctuation magnitude gives a factor of $\sim$4 in energy. 
Hence, for actually observing a scaling of $r^{-0.29 \pm 0.06}$, instead of $r^{-0.53 \pm 0.02}$, one would require that approximately 75\% of the observed EM fluctuation energy near 1~au has been generated in situ, somewhere above 14~$\rs$. 

In summary, the results from Figure~\ref{fig_dv_db_all_data_med_ajust_Emiss} can be interpreted in two ways~: 
\begin{itemize}
    \item[(A)] We may have underestimated the amplitude of the EM fluctuations by at least a factor of $\sim$2.  
    \item[(B)] There may be a significant wave generation process within the interplanetary medium. This process should slow down the radial decrease of EM fluctuation energy, and generate around 75\% of the EM  
    fluctuation energy observed near 1~au.  
    \item[(C)] The observed hypothetical magnitudes and scalings of $\dv'$ and $\db'$ do not match the intrinsic properties of $\Ewii$. 
    The present description of the solar wind is missing at least one (or more) energy contribution of a magnitude similar to that of the missing energy, $\Em$.
\end{itemize}

The computation of the excess kurtosis (Figure~\ref{fig_example_kurt_effet_std_val} and \ref{fig_kurtosis_filt_PSP_SO}) observationally attests that the amplitude of the EM 
fluctuations are estimated at a scale where most of the energy is present 
(i.e. injection scale). Moreover, our uncertainty on the determination of ($\dv,\db$) is at most of $\pm$30\% (see Section~\ref{subsec_time_average_fluct_comput} and \ref{appendix:subsec_inject_scale_filt_kurtosis}), way lower than the 200\% to 500\% variation between ($\dv,\db$) and ($\dv',\db'$), required for explaining the missing energy $\Em$.
Regarding the scenario (B), no EM 
fluctuation generation mechanism acting in such large extent in the interplanetary medium, has yet been identified to our knowledge in the current literature.
These considerations therefore suggest that scenarios (A) and (B) cannot be satisfied, and that considering scenario (C) is a legitimate option.  

%

\section{Conclusion}
\label{sec:conclusion}

This paper presents a statistical study of the solar-wind energy budget over a broad radial range. Using observations from Parker Solar Probe (PSP) and Solar Orbiter (SO), we compare the observed energetic evolution of the solar wind with the expectations derived from the large-scale non-linear ideal-MHD framework. Because electromagnetic (EM) fluctuations are commonly invoked as a source of plasma heating and solar-wind acceleration, particular attention is devoted to their energetic contribution.

\paragraph{Observations}
A particular attention is devoted to assessing the reliability of the observational dataset. Measurements from PSP and SO are sorted into radial bins defined independently for each mission, with each bin containing the same amount of data. To characterize the large-scale radial evolution, the average solar-wind properties are computed within each radial bin.

To assess the reliability of the PSP and SO datasets, we analyzed and compared the observed plasma and electromagnetic properties (bulk velocity, density, proton temperature, electron temperature, velocity fluctuations, and magnetic fluctuations) with previous studies based on similar observations, while embedding more restricted statistics or radial coverage \citep{maksimovic2020, dakeyo2022, halekas2022, Rivera2024, Rivera2025}. All analyzed plasma and electromagnetic parameters exhibit radial trends consistent with previous studies.

To our knowledge, this is the first time that the radial evolution of magnetic-field fluctuations $\db$ (expressed in Alfvén-speed units) and $\dv$, have been characterized statistically over such a near-Sun and broad radial interval, extending from 14~$\rs$ to 203~$\rs$. The observed scaling, $\db \propto r^{-0.53\pm0.02}$ (Figure~\ref{fig_dv_db_all_data_med}), remains remarkably close to the WKB dissipation free prediction $\db \propto r^{-0.5}$. That of $\dv \propto r^{-0.59\pm0.02}$ deviates a little more, decreasing slightly faster than $\db$, but remaining in similar order radial scaling.

This result places strong constraints on the amount of fluctuation energy available for dissipation and heating. Although even small departures from WKB evolution may produce substantial heating when integrated over large distances, the observations suggest that only a limited fraction of the fluctuation energy is dissipated throughout the studied radial range. Determining whether such deviations are sufficient to explain the observed plasma heating remains an important topic for future investigations.

Global invariants of the solar-wind evolution were also analyzed. The computation of the magnetic and mass fluxes further strengthens confidence in the dataset, as both quantities remain conserved on average throughout the studied radial range.

\paragraph{Energy Budget Results}

After establishing the reliability of the observational trends, we tested the large-scale stationary ideal-MHD description against observations to evaluate its ability to reproduce the energetic evolution of the solar wind. Within this framework, the total energy $\Etot$ is expected to remain constant with heliocentric distance.

Instead, we find that $\Etot$ increases by approximately $56\%$ ($\pm9\%$) between $\sim14\,\rs$ and $\sim203\,\rs$ (Figure~\ref{fig_energy_all_contrib_lin_log}). Examination of the individual contributions shows that the increase in kinetic energy $\Ec$, associated with the solar-wind acceleration, is not compensated by decreases of either the thermal or electromagnetic contributions. Consequently, $\Etot$ closely follows the radial evolution of the kinetic energy $\Ec$ and increases systematically with distance.

\section{Discussion}
\label{sec:discussion}

\paragraph{Limitations and Uncertainties}

The main identifiable sources of uncertainty in the present study are the omission of alpha particles, the absence of direct electron-temperature measurements over the SO radial range, and the fact that we neglect electron heat flux.

To verify that our conclusions do not critically depend on the omission of alpha particles, we estimated their energetic contribution using typical observed alpha-particle properties. As detailed in Appendix~\ref{appendix:sec:alpha_influence_energy_budget}, alpha particles modify the total energy budget by approximately $4\%$. Their inclusion leads to a total-energy increase of $60\%$ ($\pm9\%$), compared with $56\%$ ($\pm9\%$) without including alpha particles. 
Moreover, rather than reducing the observed imbalance, alpha particles strengthen it. This behavior arises because alpha particles contribute proportionally more to kinetic and gravitational energy than to thermal energy. Consequently, the calculation performed without alpha particles effectively represents a conservative estimate of the observed non-conservation of energy.

Regarding the electron heat flux influence, omitting it modifies the total energy budget scaling in opposite way as the alpha influence. As an estimate, considering its relative magnitude with respect to the electron heat flux $\Ethe$, and adding this last to the energy budget leads to a total energy increase of 53\% ($\pm9\%$), compared with 56\% ($\pm9\%$) without considering any influence of the electron heat flux. 
The electron flux cannot reconcile the energy budget, but it does not affect current energy interpretations. Furthermore, its influence is largely canceled out with that of the alpha particles.

Another limitation of the present study is the absence of direct electron-temperature measurements over the SO radial range. 
Electron temperatures were therefore reconstructed from the PSP observations, based on the $v$--$\Te$ relationship observed for PSP data. 
Nevertheless, the reconstructed mean profile and the associated statistical spread remain consistent both with previous observations over similar radial ranges \citep[e.g. Helios data within][]{maksimovic2020, dakeyo2022}, and show good continuity with the observations of PSP. Furthermore, Figure~\ref{fig_energy_all_contrib_lin_log} shows that the electron thermal contribution represents at most approximately $15\%$ of the total energy budget. Consequently, even substantial errors in the reconstructed electron temperatures cannot account for the observed $56\%$ ($\pm9\%$) increase in $\Etot$, since this increase is way larger than the electron contribution.

Next, a natural question is whether separating slow and fast solar-wind populations could modify the inferred energetic evolution. Such an approach is common in solar-wind studies and often provides valuable physical insight \citep{halekas2022, dakeyo2022, Halekas2023, Rivera2025, Sioulas2025, Alterman2025}.

One advantage of doing so could be identifying inter-stream energy redistribution (e.g. induced by a stream interaction region or SIR), which would allow the strength and rate of interactions to be estimated. This would be an important addition to future studies. However, quantifying the impact of SIRs will not directly help to reconcile the imbalance in the energy budget. This is because SIRs only redistributes energy and does not constitute a source of energy storage. Therefore, the energy budget must first be balanced before the role of SIRs can be investigated.

While the statistics used in this study are already extensive, some regions of the heliosphere remain sampled less densely than others. Future observations may improve the coverage of these regions and enable a more precise characterization 
of the mean trend spreads. Nevertheless, the reliability analysis presented in this work indicates that the current statistical sample is sufficient to establish the physical significance of the observed trends and that observational biases are unlikely to alter the main conclusions qualitatively.

\paragraph{Physical Interpretation}

After investigating the dominant identifiable sources of uncertainty, we conclude that the observed increase of the energy of a mean proton-electron pair $\Etot$ most likely reflects a limitation of the present large-scale stationary ideal-MHD energetic description.
These results suggest that the 
large-scale energetic formalism may neglect one or more energetically significant contributions. 

Assuming that an additional contribution $\Em$ restores exact conservation of $\Etot$, we estimated both its magnitude and radial evolution. The inferred contribution reaches approximately~:
\begin{align*}
    \Em \sim (2.0\,\pm1.2) \, (\Ethp+\Ethe+\Ewii),
\end{align*}
indicating that it is significantly larger than the combined thermal and fluctuation contributions currently considered as potential drivers of solar-wind acceleration.
Under the assumption of exact energy conservation (see Figure~\ref{fig_energy_flux_r2_Umiss}), the inferred radial evolution follows~:
\begin{align*}
    \Em \propto r^{-0.54\pm0.38},
\end{align*}
way less steep than the approximate fluctuating magnetic energy scaling of $\db^2 \propto r^{-1}$ deduced from observations. This supports that both magnitude and radial scaling of the 
missing energy differs from that of $\Ewii$.
While large scale MHD equations are well defined and follows a clear and rigorous mathematical development, applying some simplification may cause the lost of some of the physics when compared to the exact set of equations. This is where an observational budget tell us if any energetic information could has been lost within the simplification.

\paragraph{Perspectives}

The most promising avenue for future work is to revisit the assumptions underlying the present energetic formalism. If the observed imbalance reflects a genuine physical effect, then additional physical ingredients or a more complete description of the energy-flow equations may be required.
One possibility concerns the projection procedure itself. In the present framework, all energy-flow contributions are projected onto the radial direction in order to test energy conservation as a function of heliocentric distance. While this procedure is mathematically necessary, it leaves open the question of whether non-radial energy flows may contribute indirectly to the observed radial evolution.
More generally, identifying the physical origin of the inferred contribution of the missing energy $\Em$ represents an important challenge for future theoretical and observational studies. Whether it arises from unresolved physical processes, neglected terms in the energy-flow formalism, or limitations of the stationary large-scale approximation remains to be determined.

\acknowledgments{We acknowledge the NASA Parker Solar Probe Mission and the SWEAP team led by M.Stevens for use of data. We thank the instrumental Solar Wind Analyser team (SWA) for valuable discussions. This research was funded by the European Research Council ERC SLOW\_SOURCE (DLV-819189) project. This work was supported by CNRS Occitanie Ouest and LIRA. We recognise the collaborative and open nature of knowledge creation and dissemination, under the control of the academic community as expressed by Camille Noûs at http://www.cogitamus.fr/indexen.html.  Mingzhe Liu also acknowledges partial support from NASA HGIO grant 80NSSC25K7689.
}

\bibliography{bibliography}{}

@misc{DOI_MAG,
  doi = {10.48322/0YY0-BA92},
  url = {https://hpde.io/NASA/NumericalData/ParkerSolarProbe/FIELDS/MAG/Level2/RTN/FullResolution/PT0.003413S},
  author = {Bale,  Stuart D. and MacDowall,  Robert J. and Koval,  Andriy and Pulupa,  Marc and Quinn,  Timothy and Schroeder,  Peter},
  title = {PSP FIELDS Fluxgate Magnetometer (MAG) Magnetic Field Vectors,  Radial-Tangential-Normal,  RTN,  Coordinates,  Full Resolution,  Level 2 (L2),  3.413 ms Data},
  publisher = {NASA Space Physics Data Facility},
  year = {2020},
  copyright = {Creative Commons Zero v1.0 Universal}
}

@ARTICLE{bale2016,
       author = {{Bale}, S.~D. and {Goetz}, K. and {Harvey}, P.~R. and {Turin}, P. and {Bonnell}, J.~W. and {Dudok de Wit}, T. and {Ergun}, R.~E. and {MacDowall}, R.~J. and {Pulupa}, M. and {Andre}, M. and {Bolton}, M. and {Bougeret}, J. -L. and {Bowen}, T.~A. and {Burgess}, D. and {Cattell}, C.~A. and {Chandran}, B.~D.~G. and {Chaston}, C.~C. and {Chen}, C.~H.~K. and {Choi}, M.~K. and {Connerney}, J.~E. and {Cranmer}, S. and {Diaz-Aguado}, M. and {Donakowski}, W. and {Drake}, J.~F. and {Farrell}, W.~M. and {Fergeau}, P. and {Fermin}, J. and {Fischer}, J. and {Fox}, N. and {Glaser}, D. and {Goldstein}, M. and {Gordon}, D. and {Hanson}, E. and {Harris}, S.~E. and {Hayes}, L.~M. and {Hinze}, J.~J. and {Hollweg}, J.~V. and {Horbury}, T.~S. and {Howard}, R.~A. and {Hoxie}, V. and {Jannet}, G. and {Karlsson}, M. and {Kasper}, J.~C. and {Kellogg}, P.~J. and {Kien}, M. and {Klimchuk}, J.~A. and {Krasnoselskikh}, V.~V. and {Krucker}, S. and {Lynch}, J.~J. and {Maksimovic}, M. and {Malaspina}, D.~M. and {Marker}, S. and {Martin}, P. and {Martinez-Oliveros}, J. and {McCauley}, J. and {McComas}, D.~J. and {McDonald}, T. and {Meyer-Vernet}, N. and {Moncuquet}, M. and {Monson}, S.~J. and {Mozer}, F.~S. and {Murphy}, S.~D. and {Odom}, J. and {Oliverson}, R. and {Olson}, J. and {Parker}, E.~N. and {Pankow}, D. and {Phan}, T. and {Quataert}, E. and {Quinn}, T. and {Ruplin}, S.~W. and {Salem}, C. and {Seitz}, D. and {Sheppard}, D.~A. and {Siy}, A. and {Stevens}, K. and {Summers}, D. and {Szabo}, A. and {Timofeeva}, M. and {Vaivads}, A. and {Velli}, M. and {Yehle}, A. and {Werthimer}, D. and {Wygant}, J.~R.},
        title = "{The FIELDS Instrument Suite for Solar Probe Plus. Measuring the Coronal Plasma and Magnetic Field, Plasma Waves and Turbulence, and Radio Signatures of Solar Transients}",
      journal = {\ssr},
         year = 2016,
        month = dec,
       volume = {204},
       number = {1-4},
        pages = {49-82},
          doi = {10.1007/s11214-016-0244-5},
       adsurl = {https://ui.adsabs.harvard.edu/abs/2016SSRv..204...49B}
}

@article{Lopez1986solar,
       author = {{Lopez}, R.~E. and {Freeman}, J.~W.},
        title = "{Solar wind proton temperature-velocity relationship}",
      journal = {\jgr},
         year = 1986,
        month = feb,
       volume = {91},
       number = {A2},
        pages = {1701-1705},
          doi = {10.1029/JA091iA02p01701},
       adsurl = {https://ui.adsabs.harvard.edu/abs/1986JGR....91.1701L}
}

@article{Elliott2012temporal,
       author = {{Elliott}, H.~A. and {Henney}, C.~J. and {McComas}, D.~J. and {Smith}, C.~W. and {Vasquez}, B.~J.},
        title = "{Temporal and radial variation of the solar wind temperature-speed relationship}",
      journal = {Journal of Geophysical Research (Space Physics)},
         year = 2012,
        month = sep,
       volume = {117},
       number = {A9},
          eid = {A09102},
        pages = {A09102},
          doi = {10.1029/2011JA017125},
       adsurl = {https://ui.adsabs.harvard.edu/abs/2012JGRA..117.9102E}
}

@article{Schwartz1983radial,
      author = {{Schwartz}, S.~J. and {Marsch}, E.},
        title = "{The radial evolution of a single solar wind plasma parcel}",
      journal = {\jgr},
         year = 1983,
        month = dec,
       volume = {88},
       number = {A12},
        pages = {9919-9932},
          doi = {10.1029/JA088iA12p09919},
       adsurl = {https://ui.adsabs.harvard.edu/abs/1983JGR....88.9919S}
}

@ARTICLE{maksimovic2020,
       author = {{Maksimovic}, M. and {Bale}, S.~D. and {Ber{\v{c}}i{\v{c}}}, L. and {Bonnell}, J.~W. and {Case}, A.~W. and {Dudok de Wit}, T. and {Goetz}, K. and {Halekas}, J.~S. and {Harvey}, P.~R. and {Issautier}, K. and {Kasper}, J.~C. and {Korreck}, K.~E. and {Jagarlamudi}, V. Krishna and {Lahmiti}, N. and {Larson}, D.~E. and {Lecacheux}, A. and {Livi}, R. and {MacDowall}, R.~J. and {Malaspina}, D.~M. and {Martinovi{\'c}}, M.~M. and {Meyer-Vernet}, N. and {Moncuquet}, M. and {Pulupa}, M. and {Salem}, C. and {Stevens}, M.~L. and {{\v{S}}tver{\'a}k}, {\v{S}}. and {Velli}, M. and {Whittlesey}, P.~L.},
        title = "{Anticorrelation between the Bulk Speed and the Electron Temperature in the Pristine Solar Wind: First Results from the Parker Solar Probe and Comparison with Helios}",
      journal = {\apjs},
         year = 2020,
        month = feb,
       volume = {246},
       number = {2},
          eid = {62},
        pages = {62},
          doi = {10.3847/1538-4365/ab61fc},
       adsurl = {https://ui.adsabs.harvard.edu/abs/2020ApJS..246...62M}
}

@article{Kasper2015SWEAP,
      author = {{Kasper}, Justin C. and {Abiad}, Robert and {Austin}, Gerry and {Balat-Pichelin}, Marianne and {Bale}, Stuart D. and {Belcher}, John W. and {Berg}, Peter and {Bergner}, Henry and {Berthomier}, Matthieu and {Bookbinder}, Jay and {Brodu}, Etienne and {Caldwell}, David and {Case}, Anthony W. and {Chandran}, Benjamin D.~G. and {Cheimets}, Peter and {Cirtain}, Jonathan W. and {Cranmer}, Steven R. and {Curtis}, David W. and {Daigneau}, Peter and {Dalton}, Greg and {Dasgupta}, Brahmananda and {DeTomaso}, David and {Diaz-Aguado}, Millan and {Djordjevic}, Blagoje and {Donaskowski}, Bill and {Effinger}, Michael and {Florinski}, Vladimir and {Fox}, Nichola and {Freeman}, Mark and {Gallagher}, Dennis and {Gary}, S. Peter and {Gauron}, Tom and {Gates}, Richard and {Goldstein}, Melvin and {Golub}, Leon and {Gordon}, Dorothy A. and {Gurnee}, Reid and {Guth}, Giora and {Halekas}, Jasper and {Hatch}, Ken and {Heerikuisen}, Jacob and {Ho}, George and {Hu}, Qiang and {Johnson}, Greg and {Jordan}, Steven P. and {Korreck}, Kelly E. and {Larson}, Davin and {Lazarus}, Alan J. and {Li}, Gang and {Livi}, Roberto and {Ludlam}, Michael and {Maksimovic}, Milan and {McFadden}, James P. and {Marchant}, William and {Maruca}, Bennet A. and {McComas}, David J. and {Messina}, Luciana and {Mercer}, Tony and {Park}, Sang and {Peddie}, Andrew M. and {Pogorelov}, Nikolai and {Reinhart}, Matthew J. and {Richardson}, John D. and {Robinson}, Miles and {Rosen}, Irene and {Skoug}, Ruth M. and {Slagle}, Amanda and {Steinberg}, John T. and {Stevens}, Michael L. and {Szabo}, Adam and {Taylor}, Ellen R. and {Tiu}, Chris and {Turin}, Paul and {Velli}, Marco and {Webb}, Gary and {Whittlesey}, Phyllis and {Wright}, Ken and {Wu}, S.~T. and {Zank}, Gary},
        title = "{Solar Wind Electrons Alphas and Protons (SWEAP) Investigation: Design of the Solar Wind and Coronal Plasma Instrument Suite for Solar Probe Plus}",
      journal = {\ssr},
         year = 2016,
        month = dec,
       volume = {204},
       number = {1-4},
        pages = {131-186},
          doi = {10.1007/s11214-015-0206-3},
       adsurl = {https://ui.adsabs.harvard.edu/abs/2016SSRv..204..131K}
}

@ARTICLE{dakeyo2022,
       author = {{Dakeyo}, Jean-Baptiste and {Maksimovic}, Milan and {D{\'e}moulin}, Pascal and {Halekas}, Jasper and {Stevens}, Michael L.},
        title = "{Statistical Analysis of the Radial Evolution of the Solar Winds between 0.1 and 1 au and Their Semiempirical Isopoly Fluid Modeling}",
      journal = {\apj},
         year = 2022,
        month = dec,
       volume = {940},
       number = {2},
          eid = {130},
        pages = {130},
          doi = {10.3847/1538-4357/ac9b14},
archivePrefix = {arXiv},
       eprint = {2207.03898},
 primaryClass = {astro-ph.SR},
       adsurl = {https://ui.adsabs.harvard.edu/abs/2022ApJ...940..130D}
}

@ARTICLE{dakeyo2024b,
       author = {{Dakeyo}, J-B. and {Rouillard}, A.~P. and {R{\'e}ville}, V. and {D{\'e}moulin}, P. and {Maksimovic}, M. and {Chapiron}, A. and {Pinto}, R.~F. and {Louarn}, P.},
        title = "{Testing the flux tube expansion factor -- solar wind speed relation with Solar Orbiter data}",
      journal = {arXiv e-prints},
         year = 2024,
        month = aug,
          eid = {arXiv:2408.06155},
        pages = {arXiv:2408.06155},
          doi = {10.48550/arXiv.2408.06155},
archivePrefix = {arXiv},
       eprint = {2408.06155},
 primaryClass = {astro-ph.SR},
       adsurl = {https://ui.adsabs.harvard.edu/abs/2024arXiv240806155D}
}

@ARTICLE{dakeyo2026,
       author = {{Dakeyo}, Jean-Baptiste and {Ervin}, Tamar and {Bale}, Stuart and {D{\'e}moulin}, Pascal and {Sioulas}, Nikos and {R{\'e}ville}, Victor and {Liu}, Mingzhe and {Rouillard}, Alexis and {Maksimovic}, Milan and {Larson}, Davin and {Romeo}, Orlando and {Louarn}, Philippe and {Livi}, Roberto},
        title = "{On the Radial Evolution of the Solar Wind : The Source Alignment Method Applied to Parker Solar Probe and Solar Orbiter Observations}",
      journal = {arXiv e-prints},
         year = 2026,
        month = may,
          eid = {arXiv:2605.01511},
        pages = {arXiv:2605.01511},
          doi = {10.48550/arXiv.2605.01511},
archivePrefix = {arXiv},
       eprint = {2605.01511},
 primaryClass = {astro-ph.SR},
       adsurl = {https://ui.adsabs.harvard.edu/abs/2026arXiv260501511D}
}

@ARTICLE{owen_SWA2020,
       author = {{Owen}, C.~J. and {Bruno}, R. and {Livi}, S. and {Louarn}, P. and {Al Janabi}, K. and {Allegrini}, F. and {Amoros}, C. and {Baruah}, R. and {Barthe}, A. and {Berthomier}, M. and {Bordon}, S. and {Brockley-Blatt}, C. and {Brysbaert}, C. and {Capuano}, G. and {Collier}, M. and {DeMarco}, R. and {Fedorov}, A. and {Ford}, J. and {Fortunato}, V. and {Fratter}, I. and {Galvin}, A.~B. and {Hancock}, B. and {Heirtzler}, D. and {Kataria}, D. and {Kistler}, L. and {Lepri}, S.~T. and {Lewis}, G. and {Loeffler}, C. and {Marty}, W. and {Mathon}, R. and {Mayall}, A. and {Mele}, G. and {Ogasawara}, K. and {Orlandi}, M. and {Pacros}, A. and {Penou}, E. and {Persyn}, S. and {Petiot}, M. and {Phillips}, M. and {P{\v{r}}ech}, L. and {Raines}, J.~M. and {Reden}, M. and {Rouillard}, A.~P. and {Rousseau}, A. and {Rubiella}, J. and {Seran}, H. and {Spencer}, A. and {Thomas}, J.~W. and {Trevino}, J. and {Verscharen}, D. and {Wurz}, P. and {Alapide}, A. and {Amoruso}, L. and {Andr{\'e}}, N. and {Anekallu}, C. and {Arciuli}, V. and {Arnett}, K.~L. and {Ascolese}, R. and {Bancroft}, C. and {Bland}, P. and {Brysch}, M. and {Calvanese}, R. and {Castronuovo}, M. and {{\v{C}}erm{\'a}k}, I. and {Chornay}, D. and {Clemens}, S. and {Coker}, J. and {Collinson}, G. and {D'Amicis}, R. and {Dandouras}, I. and {Darnley}, R. and {Davies}, D. and {Davison}, G. and {De Los Santos}, A. and {Devoto}, P. and {Dirks}, G. and {Edlund}, E. and {Fazakerley}, A. and {Ferris}, M. and {Frost}, C. and {Fruit}, G. and {Garat}, C. and {G{\'e}not}, V. and {Gibson}, W. and {Gilbert}, J.~A. and {de Giosa}, V. and {Gradone}, S. and {Hailey}, M. and {Horbury}, T.~S. and {Hunt}, T. and {Jacquey}, C. and {Johnson}, M. and {Lavraud}, B. and {Lawrenson}, A. and {Leblanc}, F. and {Lockhart}, W. and {Maksimovic}, M. and {Malpus}, A. and {Marcucci}, F. and {Mazelle}, C. and {Monti}, F. and {Myers}, S. and {Nguyen}, T. and {Rodriguez-Pacheco}, J. and {Phillips}, I. and {Popecki}, M. and {Rees}, K. and {Rogacki}, S.~A. and {Ruane}, K. and {Rust}, D. and {Salatti}, M. and {Sauvaud}, J.~A. and {Stakhiv}, M.~O. and {Stange}, J. and {Stubbs}, T. and {Taylor}, T. and {Techer}, J. -D. and {Terrier}, G. and {Thibodeaux}, R. and {Urdiales}, C. and {Varsani}, A. and {Walsh}, A.~P. and {Watson}, G. and {Wheeler}, P. and {Willis}, G. and {Wimmer-Schweingruber}, R.~F. and {Winter}, B. and {Yardley}, J. and {Zouganelis}, I.},
        title = "{The Solar Orbiter Solar Wind Analyser (SWA) suite}",
      journal = {\aap},
         year = 2020,
        month = oct,
       volume = {642},
          eid = {A16},
        pages = {A16},
          doi = {10.1051/0004-6361/201937259},
       adsurl = {https://ui.adsabs.harvard.edu/abs/2020A&A...642A..16O}
}

@ARTICLE{kopp_holzer1976,
       author = {{Kopp}, R.~A. and {Holzer}, T.~E.},
        title = "{Dynamics of coronal hole regions. I. Steady polytropic flows with multiple critical points.}",
      journal = {\solphys},
         year = 1976,
        month = jul,
       volume = {49},
       number = {1},
        pages = {43-56},
          doi = {10.1007/BF00221484},
       adsurl = {https://ui.adsabs.harvard.edu/abs/1976SoPh...49...43K}
}

@ARTICLE{reville2020,
       author = {{R{\'e}ville}, Victor and {Velli}, Marco and {Panasenco}, Olga and {Tenerani}, Anna and {Shi}, Chen and {Badman}, Samuel T. and {Bale}, Stuart D. and {Kasper}, J.~C. and {Stevens}, Michael L. and {Korreck}, Kelly E. and {Bonnell}, J.~W. and {Case}, Anthony W. and {de Wit}, Thierry Dudok and {Goetz}, Keith and {Harvey}, Peter R. and {Larson}, Davin E. and {Livi}, Roberto and {Malaspina}, David M. and {MacDowall}, Robert J. and {Pulupa}, Marc and {Whittlesey}, Phyllis L.},
        title = "{The Role of Alfv{\'e}n Wave Dynamics on the Large-scale Properties of the Solar Wind: Comparing an MHD Simulation with Parker Solar Probe E1 Data}",
      journal = {\apjs},
         year = 2020,
        month = feb,
       volume = {246},
       number = {2},
          eid = {24},
        pages = {24},
          doi = {10.3847/1538-4365/ab4fef},
archivePrefix = {arXiv},
       eprint = {1912.03777},
 primaryClass = {astro-ph.SR},
       adsurl = {https://ui.adsabs.harvard.edu/abs/2020ApJS..246...24R}
}

@ARTICLE{halekas2022,
       author = {{Halekas}, J.~S. and {Whittlesey}, P. and {Larson}, D.~E. and {Maksimovic}, M. and {Livi}, R. and {Berthomier}, M. and {Kasper}, J.~C. and {Case}, A.~W. and {Stevens}, M.~L. and {Bale}, S.~D. and {MacDowall}, R.~J. and {Pulupa}, M.~P.},
        title = "{The Radial Evolution of the Solar Wind as Organized by Electron Distribution Parameters}",
      journal = {\apj},
         year = 2022,
        month = sep,
       volume = {936},
       number = {1},
          eid = {53},
        pages = {53},
          doi = {10.3847/1538-4357/ac85b8},
archivePrefix = {arXiv},
       eprint = {2207.06563},
 primaryClass = {astro-ph.SR},
       adsurl = {https://ui.adsabs.harvard.edu/abs/2022ApJ...936...53H}
}

@ARTICLE{fox2016,
       author = {{Fox}, N.~J. and {Velli}, M.~C. and {Bale}, S.~D. and {Decker}, R. and {Driesman}, A. and {Howard}, R.~A. and {Kasper}, J.~C. and {Kinnison}, J. and {Kusterer}, M. and {Lario}, D. and {Lockwood}, M.~K. and {McComas}, D.~J. and {Raouafi}, N.~E. and {Szabo}, A.},
        title = "{The Solar Probe Plus Mission: Humanity's First Visit to Our Star}",
      journal = {\ssr},
         year = 2016,
        month = dec,
       volume = {204},
       number = {1-4},
        pages = {7-48},
          doi = {10.1007/s11214-015-0211-6},
       adsurl = {https://ui.adsabs.harvard.edu/abs/2016SSRv..204....7F}
}

@ARTICLE{muller2020,
       author = {{M{\"u}ller}, D. and {St. Cyr}, O.~C. and {Zouganelis}, I. and {Gilbert}, H.~R. and {Marsden}, R. and {Nieves-Chinchilla}, T. and {Antonucci}, E. and {Auch{\`e}re}, F. and {Berghmans}, D. and {Horbury}, T.~S. and {Howard}, R.~A. and {Krucker}, S. and {Maksimovic}, M. and {Owen}, C.~J. and {Rochus}, P. and {Rodriguez-Pacheco}, J. and {Romoli}, M. and {Solanki}, S.~K. and {Bruno}, R. and {Carlsson}, M. and {Fludra}, A. and {Harra}, L. and {Hassler}, D.~M. and {Livi}, S. and {Louarn}, P. and {Peter}, H. and {Sch{\"u}hle}, U. and {Teriaca}, L. and {del Toro Iniesta}, J.~C. and {Wimmer-Schweingruber}, R.~F. and {Marsch}, E. and {Velli}, M. and {De Groof}, A. and {Walsh}, A. and {Williams}, D.},
        title = "{The Solar Orbiter mission. Science overview}",
      journal = {\aap},
         year = 2020,
        month = oct,
       volume = {642},
          eid = {A1},
        pages = {A1},
          doi = {10.1051/0004-6361/202038467},
archivePrefix = {arXiv},
       eprint = {2009.00861},
 primaryClass = {astro-ph.SR},
       adsurl = {https://ui.adsabs.harvard.edu/abs/2020A&A...642A...1M}
}

@ARTICLE{belcher1971,
       author = {{Belcher}, J.~W.},
        title = "{ALFV{\'E}NIC Wave Pressures and the Solar Wind}",
      journal = {\apj},
         year = 1971,
        month = sep,
       volume = {168},
        pages = {509},
          doi = {10.1086/151105},
       adsurl = {https://ui.adsabs.harvard.edu/abs/1971ApJ...168..509B}
}

@ARTICLE{chandran2009,
       author = {{Chandran}, Benjamin D.~G. and {Hollweg}, Joseph V.},
        title = "{Alfv{\'e}n Wave Reflection and Turbulent Heating in the Solar Wind from 1 Solar Radius to 1 AU: An Analytical Treatment}",
      journal = {\apj},
         year = 2009,
        month = dec,
       volume = {707},
       number = {2},
        pages = {1659-1667},
          doi = {10.1088/0004-637X/707/2/1659},
archivePrefix = {arXiv},
       eprint = {0911.1068},
 primaryClass = {astro-ph.SR},
       adsurl = {https://ui.adsabs.harvard.edu/abs/2009ApJ...707.1659C}
}

@ARTICLE{shi2023,
       author = {{Shi}, Chen and {Velli}, Marco and {Lionello}, Roberto and {Sioulas}, Nikos and {Huang}, Zesen and {Halekas}, Jasper S. and {Tenerani}, Anna and {R{\'e}ville}, Victor and {Dakeyo}, Jean-Baptiste and {Maksimovi{\'c}}, Milan and {Bale}, Stuart D.},
        title = "{Proton and Electron Temperatures in the Solar Wind and Their Correlations with the Solar Wind Speed}",
      journal = {\apj},
         year = 2023,
        month = feb,
       volume = {944},
       number = {1},
          eid = {82},
        pages = {82},
          doi = {10.3847/1538-4357/acb341},
archivePrefix = {arXiv},
       eprint = {2301.00852},
 primaryClass = {astro-ph.SR},
       adsurl = {https://ui.adsabs.harvard.edu/abs/2023ApJ...944...82S}
}

@ARTICLE{Horbury_MAG2020,
       author = {{Horbury}, T.~S. and {O'Brien}, H. and {Carrasco Blazquez}, I. and {Bendyk}, M. and {Brown}, P. and {Hudson}, R. and {Evans}, V. and {Oddy}, T.~M. and {Carr}, C.~M. and {Beek}, T.~J. and {Cupido}, E. and {Bhattacharya}, S. and {Dominguez}, J. -A. and {Matthews}, L. and {Myklebust}, V.~R. and {Whiteside}, B. and {Bale}, S.~D. and {Baumjohann}, W. and {Burgess}, D. and {Carbone}, V. and {Cargill}, P. and {Eastwood}, J. and {Erd{\"o}s}, G. and {Fletcher}, L. and {Forsyth}, R. and {Giacalone}, J. and {Glassmeier}, K. -H. and {Goldstein}, M.~L. and {Hoeksema}, T. and {Lockwood}, M. and {Magnes}, W. and {Maksimovic}, M. and {Marsch}, E. and {Matthaeus}, W.~H. and {Murphy}, N. and {Nakariakov}, V.~M. and {Owen}, C.~J. and {Owens}, M. and {Rodriguez-Pacheco}, J. and {Richter}, I. and {Riley}, P. and {Russell}, C.~T. and {Schwartz}, S. and {Vainio}, R. and {Velli}, M. and {Vennerstrom}, S. and {Walsh}, R. and {Wimmer-Schweingruber}, R.~F. and {Zank}, G. and {M{\"u}ller}, D. and {Zouganelis}, I. and {Walsh}, A.~P.},
        title = "{The Solar Orbiter magnetometer}",
      journal = {\aap},
         year = 2020,
        month = oct,
       volume = {642},
          eid = {A9},
        pages = {A9},
          doi = {10.1051/0004-6361/201937257},
       adsurl = {https://ui.adsabs.harvard.edu/abs/2020A&A...642A...9H}
}

@ARTICLE{Bale_2023,
       author = {{Bale}, S.~D. and {Drake}, J.~F. and {McManus}, M.~D. and {Desai}, M.~I. and {Badman}, S.~T. and {Larson}, D.~E. and {Swisdak}, M. and {Horbury}, T.~S. and {Raouafi}, N.~E. and {Phan}, T. and {Velli}, M. and {McComas}, D.~J. and {Cohen}, C.~M.~S. and {Mitchell}, D. and {Panasenco}, O. and {Kasper}, J.~C.},
        title = "{Interchange reconnection as the source of the fast solar wind within coronal holes}",
      journal = {\nat},
         year = 2023,
        month = jun,
       volume = {618},
       number = {7964},
        pages = {252-256},
          doi = {10.1038/s41586-023-05955-3},
archivePrefix = {arXiv},
       eprint = {2208.07932},
 primaryClass = {astro-ph.SR},
       adsurl = {https://ui.adsabs.harvard.edu/abs/2023Natur.618..252B}
}

@ARTICLE{Sioulas2025,
       author = {{Sioulas}, Nikos and {Huang}, Zesen and {Shi}, Chen and {Velli}, Marco and {Tenerani}, Anna and {Bowen}, Trevor A. and {Bale}, Stuart D. and {Huang}, Jia and {Vlahos}, Loukas and {Woodham}, L.~D. and {Horbury}, T.~S. and {de Wit}, Thierry Dudok and {Larson}, Davin and {Kasper}, Justin and {Owen}, Christopher J. and {Stevens}, Michael L. and {Case}, Anthony and {Pulupa}, Marc and {Malaspina}, David M. and {Bonnell}, J.~W. and {Livi}, Roberto and {Goetz}, Keith and {Harvey}, Peter R. and {MacDowall}, Robert J. and {Maksimovi{\'c}}, Milan and {Louarn}, P. and {Fedorov}, A.},
        title = "{Magnetic Field Spectral Evolution in the Inner Heliosphere}",
      journal = {\apjl},
         year = 2025,
        month = jan,
       volume = {943},
       number = {1},
          eid = {L8},
        pages = {L8},
          doi = {10.3847/2041-8213/acaeff},
archivePrefix = {arXiv},
       eprint = {2209.02451},
 primaryClass = {astro-ph.SR},
       adsurl = {https://ui.adsabs.harvard.edu/abs/2023ApJ...943L...8S}
}

@ARTICLE{Alazraki1971,
       author = {{Alazraki}, G. and {Couturier}, P.},
        title = "{Solar Wind Accejeration Caused by the Gradient of Alfven Wave Pressure}",
      journal = {\aap},
         year = 1971,
        month = aug,
       volume = {13},
        pages = {380},
       adsurl = {https://ui.adsabs.harvard.edu/abs/1971A&A....13..380A}
}

@ARTICLE{Rivera2024,
       author = {{Rivera}, Yeimy J. and {Badman}, Samuel T. and {Stevens}, Michael L. and {Verniero}, Jaye L. and {Stawarz}, Julia E. and {Shi}, Chen and {Raines}, Jim M. and {Paulson}, Kristoff W. and {Owen}, Christopher J. and {Niembro}, Tatiana and {Louarn}, Philippe and {Livi}, Stefano A. and {Lepri}, Susan T. and {Kasper}, Justin C. and {Horbury}, Timothy S. and {Halekas}, Jasper S. and {Dewey}, Ryan M. and {De Marco}, Rossana and {Bale}, Stuart D.},
        title = "{In situ observations of large-amplitude Alfv{\'e}n waves heating and accelerating the solar wind}",
      journal = {Science},
         year = 2024,
        month = aug,
       volume = {385},
       number = {6712},
        pages = {962-966},
          doi = {10.1126/science.adk6953},
archivePrefix = {arXiv},
       eprint = {2409.00267},
 primaryClass = {astro-ph.SR},
       adsurl = {https://ui.adsabs.harvard.edu/abs/2024Sci...385..962R}
}

@ARTICLE{Rivera2025,
       author = {{Rivera}, Yeimy J. and {Badman}, Samuel T. and {Verniero}, J.~L. and {Varesano}, Tania and {Stevens}, Michael L. and {Stawarz}, Julia E. and {Reeves}, Katharine K. and {Raines}, Jim M. and {Raymond}, John C. and {Owen}, Christopher J. and {Livi}, Stefano A. and {Lepri}, Susan T. and {Landi}, Enrico and {Halekas}, Jasper. S. and {Ervin}, Tamar and {Dewey}, Ryan M. and {De Marco}, Rossana and {D'Amicis}, Raffaella and {Dakeyo}, Jean-Baptiste and {Bale}, Stuart D. and {Alterman}, B.~L.},
        title = "{Differentiating the Acceleration Mechanisms in the Slow and Alfv{\'e}nic Slow Solar Wind}",
      journal = {\apj},
         year = 2025,
        month = feb,
       volume = {980},
       number = {1},
          eid = {70},
        pages = {70},
          doi = {10.3847/1538-4357/ada699},
archivePrefix = {arXiv},
       eprint = {2501.02163},
 primaryClass = {astro-ph.SR},
       adsurl = {https://ui.adsabs.harvard.edu/abs/2025ApJ...980...70R}
}

@ARTICLE{Ervin2024b,
       author = {{Ervin}, Tamar and {Bale}, Stuart D. and {Badman}, Samuel T. and {Bowen}, Trevor A. and {Riley}, Pete and {Paulson}, Kristoff and {Rivera}, Yeimy J. and {Romeo}, Orlando and {Sioulas}, Nikos and {Larson}, Davin and {Verniero}, Jaye L. and {Dewey}, Ryan M. and {Huang}, Jia},
        title = "{Near Subsonic Solar Wind Outflow from an Active Region}",
      journal = {\apj},
         year = 2024,
        month = sep,
       volume = {972},
       number = {1},
          eid = {129},
        pages = {129},
          doi = {10.3847/1538-4357/ad57c4},
archivePrefix = {arXiv},
       eprint = {2405.15844},
 primaryClass = {astro-ph.SR},
       adsurl = {https://ui.adsabs.harvard.edu/abs/2024ApJ...972..129E}
}

@ARTICLE{Bale2005,
       author = {{Bale}, S.~D. and {Kellogg}, P.~J. and {Mozer}, F.~S. and {Horbury}, T.~S. and {Reme}, H.},
        title = "{Measurement of the Electric Fluctuation Spectrum of Magnetohydrodynamic Turbulence}",
      journal = {\prl},
         year = 2005,
        month = jun,
       volume = {94},
       number = {21},
          eid = {215002},
        pages = {215002},
          doi = {10.1103/PhysRevLett.94.215002},
archivePrefix = {arXiv},
       eprint = {physics/0503103},
 primaryClass = {physics.space-ph},
       adsurl = {https://ui.adsabs.harvard.edu/abs/2005PhRvL..94u5002B}
}

@ARTICLE{Huang2023,
       author = {{Huang}, Zesen and {Sioulas}, Nikos and {Shi}, Chen and {Velli}, Marco and {Bowen}, Trevor and {Davis}, Nooshin and {Chandran}, B.~D.~G. and {Matteini}, Lorenzo and {Kang}, Ning and {Shi}, Xiaofei and {Huang}, Jia and {Bale}, Stuart D. and {Kasper}, J.~C. and {Larson}, Davin E. and {Livi}, Roberto and {Whittlesey}, P.~L. and {Rahmati}, Ali and {Paulson}, Kristoff and {Stevens}, M. and {Case}, A.~W. and {de Wit}, Thierry Dudok and {Malaspina}, David M. and {Bonnell}, J.~W. and {Goetz}, Keith and {Harvey}, Peter R. and {MacDowall}, Robert J.},
        title = "{New Observations of Solar Wind 1/f Turbulence Spectrum from Parker Solar Probe}",
      journal = {\apjl},
         year = 2023,
        month = jun,
       volume = {950},
       number = {1},
          eid = {L8},
        pages = {L8},
          doi = {10.3847/2041-8213/acd7f2},
archivePrefix = {arXiv},
       eprint = {2303.00843},
 primaryClass = {astro-ph.SR},
       adsurl = {https://ui.adsabs.harvard.edu/abs/2023ApJ...950L...8H}
}

@ARTICLE{Alexandrova2013,
       author = {{Alexandrova}, O. and {Chen}, C.~H.~K. and {Sorriso-Valvo}, L. and {Horbury}, T.~S. and {Bale}, S.~D.},
        title = "{Solar Wind Turbulence and the Role of Ion Instabilities}",
      journal = {\ssr},
         year = 2013,
        month = oct,
       volume = {178},
       number = {2-4},
        pages = {101-139},
          doi = {10.1007/s11214-013-0004-8},
archivePrefix = {arXiv},
       eprint = {1306.5336},
 primaryClass = {astro-ph.SR},
       adsurl = {https://ui.adsabs.harvard.edu/abs/2013SSRv..178..101A}
}

@ARTICLE{Sioulas2023,
       author = {{Sioulas}, Nikos and {Huang}, Zesen and {Shi}, Chen and {Velli}, Marco and {Tenerani}, Anna and {Bowen}, Trevor A. and {Bale}, Stuart D. and {Huang}, Jia and {Vlahos}, Loukas and {Woodham}, L.~D. and {Horbury}, T.~S. and {de Wit}, Thierry Dudok and {Larson}, Davin and {Kasper}, Justin and {Owen}, Christopher J. and {Stevens}, Michael L. and {Case}, Anthony and {Pulupa}, Marc and {Malaspina}, David M. and {Bonnell}, J.~W. and {Livi}, Roberto and {Goetz}, Keith and {Harvey}, Peter R. and {MacDowall}, Robert J. and {Maksimovi{\'c}}, Milan and {Louarn}, P. and {Fedorov}, A.},
        title = "{Magnetic Field Spectral Evolution in the Inner Heliosphere}",
      journal = {\apjl},
         year = 2023,
        month = jan,
       volume = {943},
       number = {1},
          eid = {L8},
        pages = {L8},
          doi = {10.3847/2041-8213/acaeff},
archivePrefix = {arXiv},
       eprint = {2209.02451},
 primaryClass = {astro-ph.SR},
       adsurl = {https://ui.adsabs.harvard.edu/abs/2023ApJ...943L...8S}
}

@ARTICLE{Huang2025,
       author = {{Huang}, Zesen and {Velli}, Marco and {Chandran}, B.~D.~G. and {Shi}, Chen and {Ding}, Yuliang and {Matteini}, Lorenzo and {Choi}, Kyung-Eun},
        title = "{Two Types of $1/f$ Range in Solar Wind Turbulence}",
      journal = {arXiv e-prints},
         year = 2025,
        month = jun,
          eid = {arXiv:2506.17523},
        pages = {arXiv:2506.17523},
          doi = {10.48550/arXiv.2506.17523},
archivePrefix = {arXiv},
       eprint = {2506.17523},
 primaryClass = {astro-ph.SR},
       adsurl = {https://ui.adsabs.harvard.edu/abs/2025arXiv250617523H}
}

@ARTICLE{QTN_ref_2020,
       author = {{Moncuquet}, Michel and {Meyer-Vernet}, Nicole and {Issautier}, Karine and {Pulupa}, Marc and {Bonnell}, J.~W. and {Bale}, Stuart D. and {Dudok de Wit}, Thierry and {Goetz}, Keith and {Griton}, L{\'e}a and {Harvey}, Peter R. and {MacDowall}, Robert J. and {Maksimovic}, Milan and {Malaspina}, David M.},
        title = "{First In Situ Measurements of Electron Density and Temperature from Quasi-thermal Noise Spectroscopy with Parker Solar Probe/FIELDS}",
      journal = {\apjs},
         year = 2020,
        month = feb,
       volume = {246},
       number = {2},
          eid = {44},
        pages = {44},
          doi = {10.3847/1538-4365/ab5a84},
archivePrefix = {arXiv},
       eprint = {1912.02518},
 primaryClass = {astro-ph.SR},
       adsurl = {https://ui.adsabs.harvard.edu/abs/2020ApJS..246...44M}
}

@misc{Moestl2020_icme_catalog,
    author    = {{Moestl}, C. and {Davies}, E. and {Weiler}, E.},
    title     = "{HELIO4CAST Interplanetary Coronal Mass Ejection Catalog v2.3}",
    year = "2020",
    url = "https://figshare.com/articles/dataset/HELCATS_Interplanetary_Coronal_Mass_Ejection_Catalog_v2_0/6356420",
     doi = "10.6084/m9.figshare.6356420.v23"
}

@ARTICLE{Halekas2023,
       author = {{Halekas}, J.~S. and {Bale}, S.~D. and {Berthomier}, M. and {Chandran}, B.~D.~G. and {Drake}, J.~F. and {Kasper}, J.~C. and {Klein}, K.~G. and {Larson}, D.~E. and {Livi}, R. and {Pulupa}, M.~P. and {Stevens}, M.~L. and {Verniero}, J.~L. and {Whittlesey}, P.},
        title = "{Quantifying the Energy Budget in the Solar Wind from 13.3 to 100 Solar Radii}",
      journal = {\apj},
         year = 2023,
        month = jul,
       volume = {952},
       number = {1},
          eid = {26},
        pages = {26},
          doi = {10.3847/1538-4357/acd769},
archivePrefix = {arXiv},
       eprint = {2305.13424},
 primaryClass = {astro-ph.SR},
       adsurl = {https://ui.adsabs.harvard.edu/abs/2023ApJ...952...26H}
}

@ARTICLE{SPAN_E_ref_2020,
       author = {{Whittlesey}, Phyllis L. and {Larson}, Davin E. and {Kasper}, Justin C. and {Halekas}, Jasper and {Abatcha}, Mamuda and {Abiad}, Robert and {Berthomier}, M. and {Case}, A.~W. and {Chen}, Jianxin and {Curtis}, David W. and {Dalton}, Gregory and {Klein}, Kristopher G. and {Korreck}, Kelly E. and {Livi}, Roberto and {Ludlam}, Michael and {Marckwordt}, Mario and {Rahmati}, Ali and {Robinson}, Miles and {Slagle}, Amanda and {Stevens}, M.~L. and {Tiu}, Chris and {Verniero}, J.~L.},
        title = "{The Solar Probe ANalyzers{\textemdash}Electrons on the Parker Solar Probe}",
      journal = {\apjs},
         year = 2020,
        month = feb,
       volume = {246},
       number = {2},
          eid = {74},
        pages = {74},
          doi = {10.3847/1538-4365/ab7370},
archivePrefix = {arXiv},
       eprint = {2002.04080},
 primaryClass = {astro-ph.IM},
       adsurl = {https://ui.adsabs.harvard.edu/abs/2020ApJS..246...74W}
}

@ARTICLE{2021LiuAA,
       author = {{Liu}, M. and {Issautier}, K. and {Meyer-Vernet}, N. and {Moncuquet}, M. and {Maksimovic}, M. and {Halekas}, J.~S. and {Huang}, J. and {Griton}, L. and {Bale}, S. and {Bonnell}, J.~W. and {Case}, A.~W. and {Goetz}, K. and {Harvey}, P.~R. and {Kasper}, J.~C. and {MacDowall}, R.~J. and {Malaspina}, D.~M. and {Pulupa}, M. and {Stevens}, M.~L.},
        title = "{Solar wind energy flux observations in the inner heliosphere: first results from Parker Solar Probe}",
      journal = {\aap},
         year = 2021,
        month = jun,
       volume = {650},
          eid = {A14},
        pages = {A14},
          doi = {10.1051/0004-6361/202039615},
archivePrefix = {arXiv},
       eprint = {2101.03121},
 primaryClass = {astro-ph.SR},
       adsurl = {https://ui.adsabs.harvard.edu/abs/2021A&A...650A..14L}
}

@ARTICLE{2023LiuAA,
       author = {{Liu}, M. and {Issautier}, K. and {Moncuquet}, M. and {Meyer-Vernet}, N. and {Maksimovic}, M. and {Huang}, J. and {Martinovic}, M.~M. and {Griton}, L. and {Chrysaphi}, N. and {Jagarlamudi}, V.~K. and {Bale}, S.~D. and {Pulupa}, M. and {Kasper}, J.~C. and {Stevens}, M.~L.},
        title = "{Total electron temperature derived from quasi-thermal noise spectroscopy in the pristine solar wind from Parker Solar Probe observations}",
      journal = {\aap},
         year = 2023,
        month = jun,
       volume = {674},
          eid = {A49},
        pages = {A49},
          doi = {10.1051/0004-6361/202245450},
archivePrefix = {arXiv},
       eprint = {2303.11035},
 primaryClass = {astro-ph.SR},
       adsurl = {https://ui.adsabs.harvard.edu/abs/2023A&A...674A..49L}
}

@ARTICLE{Sorriso1999,
       author = {{Sorriso-Valvo}, Luca and {Carbone}, Vincenzo and {Veltri}, Pierluigi and {Consolini}, Giuseppe and {Bruno}, Roberto},
        title = "{Intermittency in the solar wind turbulence through probability distribution functions of fluctuations}",
      journal = {\grl},
         year = 1999,
        month = jul,
       volume = {26},
       number = {13},
        pages = {1801-1804},
          doi = {10.1029/1999GL900270},
archivePrefix = {arXiv},
       eprint = {physics/9903043},
 primaryClass = {physics.plasm-ph},
       adsurl = {https://ui.adsabs.harvard.edu/abs/1999GeoRL..26.1801S}
}

@ARTICLE{Peng2024,
       author = {{Peng}, Jingyu and {He}, Jiansen and {Duan}, Die and {Verscharen}, Daniel},
        title = "{Observations of Preferential Heating and Acceleration of {\ensuremath{\alpha}}-particles in the Young Solar Wind by Parker Solar Probe}",
      journal = {\apj},
         year = 2024,
        month = dec,
       volume = {977},
       number = {1},
          eid = {27},
        pages = {27},
          doi = {10.3847/1538-4357/ad79fa},
       adsurl = {https://ui.adsabs.harvard.edu/abs/2024ApJ...977...27P}
}

@ARTICLE{Coleman1968,
       author = {{Coleman}, Jr., Paul J.},
        title = "{Turbulence, Viscosity, and Dissipation in the Solar-Wind Plasma}",
      journal = {\apj},
         year = 1968,
        month = aug,
       volume = {153},
        pages = {371},
          doi = {10.1086/149674},
       adsurl = {https://ui.adsabs.harvard.edu/abs/1968ApJ...153..371C}
}

@ARTICLE{Kasper2019,
       author = {{Kasper}, J.~C. and {Bale}, S.~D. and {Belcher}, J.~W. and {Berthomier}, M. and {Case}, A.~W. and {Chandran}, B.~D.~G. and {Curtis}, D.~W. and {Gallagher}, D. and {Gary}, S.~P. and {Golub}, L. and {Halekas}, J.~S. and {Ho}, G.~C. and {Horbury}, T.~S. and {Hu}, Q. and {Huang}, J. and {Klein}, K.~G. and {Korreck}, K.~E. and {Larson}, D.~E. and {Livi}, R. and {Maruca}, B. and {Lavraud}, B. and {Louarn}, P. and {Maksimovic}, M. and {Martinovic}, M. and {McGinnis}, D. and {Pogorelov}, N.~V. and {Richardson}, J.~D. and {Skoug}, R.~M. and {Steinberg}, J.~T. and {Stevens}, M.~L. and {Szabo}, A. and {Velli}, M. and {Whittlesey}, P.~L. and {Wright}, K.~H. and {Zank}, G.~P. and {MacDowall}, R.~J. and {McComas}, D.~J. and {McNutt}, R.~L. and {Pulupa}, M. and {Raouafi}, N.~E. and {Schwadron}, N.~A.},
        title = "{Alfv{\'e}nic velocity spikes and rotational flows in the near-Sun solar wind}",
      journal = {\nat},
         year = 2019,
        month = dec,
       volume = {576},
       number = {7786},
        pages = {228-231},
          doi = {10.1038/s41586-019-1813-z},
       adsurl = {https://ui.adsabs.harvard.edu/abs/2019Natur.576..228K}
}

@ARTICLE{Chandran2025,
       author = {{Chandran}, Benjamin Divakar Giles and {Adkins}, Toby and {Bale}, Stuart D. and {David}, Vincent and {Halekas}, Jasper and {Klein}, Kristopher and {Meyrand}, Romain and {Perez}, Jean C. and {Shoda}, Munehito and {Squire}, Jonathan and {Yerger}, Evan Lowell},
        title = "{A two-fluid solar-wind model with intermittent Alfv{\'e}nic turbulence}",
      journal = {Journal of Plasma Physics},
         year = 2025,
        month = aug,
       volume = {91},
       number = {4},
          eid = {E125},
        pages = {E125},
          doi = {10.1017/S0022377825100640},
       adsurl = {https://ui.adsabs.harvard.edu/abs/2025JPlPh..91E.125C}
}

@ARTICLE{Bazer1963,
       author = {{Bazer}, J. and {Hurley}, J.},
        title = "{Geometrical Hydromagnetics}",
      journal = {\jgr},
         year = 1963,
        month = jan,
       volume = {68},
       number = {1},
        pages = {147-174},
          doi = {10.1029/JZ068i001p00147},
       adsurl = {https://ui.adsabs.harvard.edu/abs/1963JGR....68..147B}
}

@ARTICLE{Webb2024,
       author = {{Webb}, Gary M. and {Anco}, Stephen C. and {Meleshko}, Sergey V. and {Kaptsov}, Evgeniy I.},
        title = "{Noether's theorems and conservation laws in magnetohydrodynamics and Chew-Goldberger-Low plasmas}",
      journal = {Reviews of Modern Plasma Physics},
         year = 2024,
        month = nov,
       volume = {8},
       number = {1},
          eid = {33},
        pages = {33},
          doi = {10.1007/s41614-024-00168-1},
       adsurl = {https://ui.adsabs.harvard.edu/abs/2024RvMPP...8...33W}
}

@book{Goedbloed2004, place={Cambridge}, title={Principles of Magnetohydrodynamics: With Applications to Laboratory and Astrophysical Plasmas}, publisher={Cambridge University Press}, author={Goedbloed, J. P. Hans and Poedts, Stefaan}, year={2004}}

@ARTICLE{Alterman2025,
       author = {{Alterman}, B.~L.},
        title = "{Characterizing the Impact of Alfv{\'e}n Wave Forcing in Interplanetary Space on the Distribution of Near-Earth Solar Wind Speeds}",
      journal = {\apjl},
         year = 2025,
        month = may,
       volume = {984},
       number = {2},
          eid = {L64},
        pages = {L64},
          doi = {10.3847/2041-8213/add0a6},
archivePrefix = {arXiv},
       eprint = {2504.18350},
 primaryClass = {astro-ph.SR},
       adsurl = {https://ui.adsabs.harvard.edu/abs/2025ApJ...984L..64A}
}

@ARTICLE{Ulysses_mission_ref,
       author = {{Wenzel}, K.~P. and {Marsden}, R.~G. and {Page}, D.~E. and {Smith}, E.~J.},
        title = "{The ULYSSES Mission}",
      journal = {\aaps},
         year = 1992,
        month = jan,
       volume = {92},
        pages = {207},
       adsurl = {https://ui.adsabs.harvard.edu/abs/1992A&AS...92..207W}
}

@ARTICLE{SWOOP_ref,
       author = {{Bame}, S.~J. and {McComas}, D.~J. and {Barraclough}, B.~L. and {Phillips}, J.~L. and {Sofaly}, K.~J. and {Chavez}, J.~C. and {Goldstein}, B.~E. and {Sakurai}, R.~K.},
        title = "{The ULYSSES solar wind plasma experiment}",
      journal = {\aaps},
         year = 1992,
        month = jan,
       volume = {92},
       number = {2},
        pages = {237-265},
       adsurl = {https://ui.adsabs.harvard.edu/abs/1992A&AS...92..237B}
}

@ARTICLE{Verscharen2019,
       author = {{Verscharen}, Daniel and {Klein}, Kristopher G. and {Maruca}, Bennett A.},
        title = "{The multi-scale nature of the solar wind}",
      journal = {Living Reviews in Solar Physics},
         year = 2019,
        month = dec,
       volume = {16},
       number = {1},
          eid = {5},
        pages = {5},
          doi = {10.1007/s41116-019-0021-0},
archivePrefix = {arXiv},
       eprint = {1902.03448},
 primaryClass = {physics.space-ph},
       adsurl = {https://ui.adsabs.harvard.edu/abs/2019LRSP...16....5V}
}

@ARTICLE{Livi2022_SPI_ref,
       author = {{Livi}, Roberto and {Larson}, Davin E. and {Kasper}, Justin C. and {Abiad}, Robert and {Case}, A.~W. and {Klein}, Kristopher G. and {Curtis}, David W. and {Dalton}, Gregory and {Stevens}, Michael and {Korreck}, Kelly E. and {Ho}, George and {Robinson}, Miles and {Tiu}, Chris and {Whittlesey}, Phyllis L. and {Verniero}, Jaye L. and {Halekas}, Jasper and {McFadden}, James and {Marckwordt}, Mario and {Slagle}, Amanda and {Abatcha}, Mamuda and {Rahmati}, Ali and {McManus}, Michael D.},
        title = "{The Solar Probe ANalyzer-Ions on the Parker Solar Probe}",
      journal = {\apj},
         year = 2022,
        month = oct,
       volume = {938},
       number = {2},
          eid = {138},
        pages = {138},
          doi = {10.3847/1538-4357/ac93f5},
       adsurl = {https://ui.adsabs.harvard.edu/abs/2022ApJ...938..138L}
}
\bibliographystyle{aasjournal}

\appendix
\newcommand{\mal}{m_\mathrm{\alpha}}
\newcommand{\val}{v_\mathrm{\alpha}}
\newcommand{\dval}{\delta v_\mathrm{\alpha}}
\newcommand{\Tal}{T_\mathrm{\alpha}}
\newcommand{\nal}{n_\mathrm{\alpha}}
\newcommand{\Aal}{A_\mathrm{\alpha}}
\newcommand{\Ethal}{E_\mathrm{th}^{(\alpha)}}

\section{Data}
\label{appendix:sec_data}

\subsection{Observations Pre-processing}
\label{appendix:sec_data_preprocessing}

The field of view (FOV) of the SPAN-I instrument is affected by the thermal shield, as well as by the instrument orientation and the spacecraft velocity relative to the solar wind bulk flow. As a result, the velocity distribution function (VDF) is not always fully contained within the instrument FOV. Time intervals for which the VDF is significantly truncated are therefore removed. 
\\
This occurs when the proton density $n_p$ is lower than 10\% of its daily average value (i.e. 90\% relative decrease). 
The total proton temperature measured by SPAN-I is adjusted to better match the radial temperature measured by the Solar Probe Cup (SPC), which has a more favorable FOV in the radial direction. A previous comparison between SPAN-I and SPC measurements showed that the radial temperature component satisfies $T_{zz} = T_{r|\spani} \approx 2 \: T_{r|\spc}$ \citep{dakeyo2022}. \\
We therefore correct the SPAN-I radial temperature by applying a factor of $1/2$ to ensure consistency with SPC measurements. 
This leads to the following expression for the total proton temperature~: 
\begin{align}
    T_{\spani} = \frac{T_{xx} + T_{yy} + T_{zz}/2}{3} .
\end{align}

The instrumental uncertainties for PSP data are approximately 3\% for the bulk speed and 10--15\% for the total proton temperature from SPAN-I; 10\% for the density and $\sim$20\% for the electron temperature derived from QTN observations of FIELDS \citep{QTN_ref_2020,2023LiuAA,2021LiuAA}; and 10--15\% for the electron temperature from SPAN-E (private communication with the SPAN-E team).
Instrumental uncertainties for SO are not explicitly provided. However, given that the FOV of PAS is more favorable than that of SPAN-I, and assuming similar performance for the MAG and FIELDS instruments, we adopt PSP instrumental uncertainties as a proxy for SO observations.\\
Interplanetary coronal mass ejections (ICMEs) are removed using the ICMECAT catalog \citep{Moestl2020_icme_catalog} for both PSP and SO datasets, and by applying the ICME identification criteria of \cite{Elliott2012temporal} to SO data, following the approach of \cite{dakeyo2024b}. Data are excluded when at least one of the following conditions on the plasma $\beta$ and proton temperature $\Tp$ is satisfied:
$(i)$ $\beta < 0.1$,  
$(ii)$ $\Tp / T_{ex} < 0.5$,  
where $T_{ex}$ is the expected temperature given by the scaling law  
$T_{ex}~=~486.5 \times \vo - 1.2476\times 10^5K$ from \cite{Lopez1986solar}, and $\vo$ the wind bulk speed. %
ICMEs are assumed to have a minimum duration of 6 hours. In addition, we remove data within 24~hours prior to and 15~hours following each detected ICME. Finally, wind intervals with speeds exceeding 800~km/s are also excluded, as they are considered likely ICME-related.

All these criteria aim to minimize the influence of the ICMEs event on solar wind data statistics. We cannot guarantee that all ICME-like events are identified and removed. However, we expect the combination of criteria from \cite[originally applied for HELIOS and assumed to be adapted as well for SO]{Elliott2012temporal} and the ICMECAT catalog to remove most of them. While it is possible that ICME-like events could remain in the filtered dataset, the large sample size aims to compensate for this. This helps to minimize the impact of ICMEs on the estimation of solar wind mean properties.

\subsection{Injection Scale Data Filtering from Excess Kurtosis}
\label{appendix:subsec_inject_scale_filt_kurtosis}

The excess kurtosis computed for the fluctuations allow us to determine whether the assumption made considering that we compute properties nearby the injection scales are correct or not. 
EM fluctuations relates to both $\dv$ and $\dB$. Rather than computing the kurtosis of each of these quantity, one can compute the kurtosis of the electric field distribution, which under the current Ideal Ohm's law assumption, embed both magnetic and velocity fluctuations effect. 
Considering so, the excess kurtosis is estimated for each component of the electric field $\Evect = - \vvect \times \Bvect$ as~:
\begin{align}
    \kappa_j 
    = \kurt(E_j)
    = \Bigg \langle \bigg( \frac{E - \mu}{\sigma} \bigg)^4 \Bigg \rangle - 3
    \qquad \text{and} \qquad
    \kappa = \frac{\sum_j \kappa_j }{3}
    \label{eq:express_average_kurtosis}
\end{align}
where $\mu$ is the average of the data sample, $\sigma$ is its standard deviation, $\langle . \rangle$ defines an average over the entire data sample, and $j = \{ R,T,N\}$. 
The average excess kurtosis $\kappa$, is then used to defined whether the EM fluctuations can be considered nearly Gaussian or not. 

\begin{figure*}[ht]
    \centering
    \includegraphics[width=12.5cm]{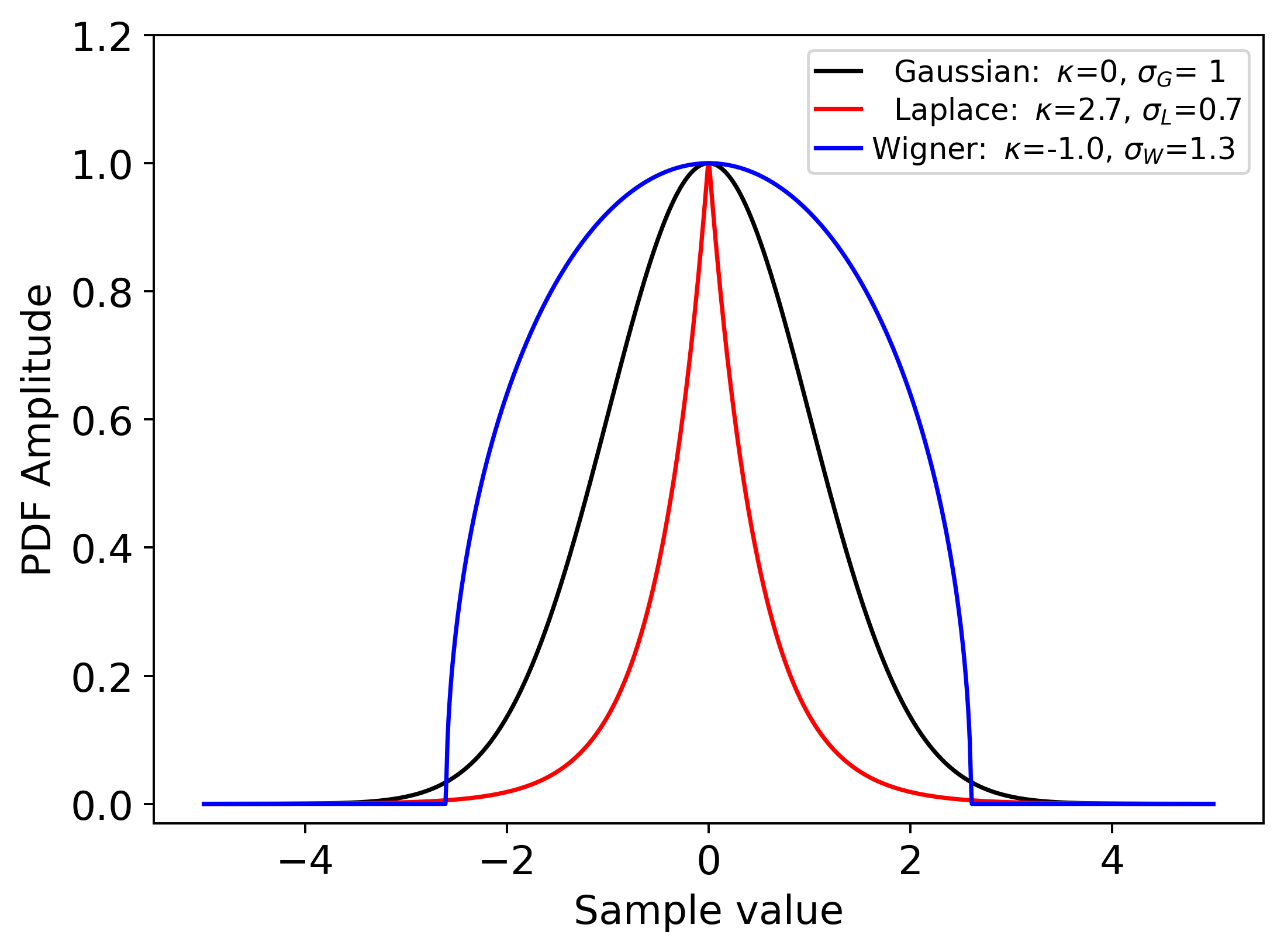}
    \caption{Probability distribution function (PDF) of a Gaussian (black), Laplace (red), and Wigner (blue) distributions. The associated excess kurtosis $\kappa$, and standard deviation $\sigma$ are indicated in legend. The different parameters are~: amplitude (A), standard deviation ($\sigma$), mean (m), radius (R), scale parameter ($\mu$). The parameter values are~: Gaussian ($A = 1$, $m = 0$, $\sigma=1$); Laplace ($A=1$, $m=0$, $\mu=0.5$); Wigner ($R=2.6$). Wigner distribution is rescaled to have the same amplitude as the Gaussian distribution of reference. This has no impact neither on the kurtosis nor on the standard deviation value.
    }
    \label{fig_example_kurt_effet_std_val}
\end{figure*}

To define a level of confidence, we estimate the excess kurtosis values 
that corresponds to a variation of 30\% of the standard deviation from a Gaussian. Figure~\ref{fig_example_kurt_effet_std_val} shows the distributions considered for defining the minimal and maximal accepted kurtosis values. 
Gaussian, Laplace and Wigner semi-circle distributions are respectively taken as reference kurtosis case, high kurtosis case and low kurtosis case~:
\begin{align}
    \text{Gaussian} : \qquad G(x) &=  A \times e^{-(x-m)^2/(2\sigma^2)} \\
    \text{Laplace}: \qquad L(x) &= A/(2\mu) \times e^{-|x-m|/\mu}\\
    \text{Wigner}: \qquad W(x) &= 2/(\pi R^2) \times \sqrt{R^2 - x^2},
\end{align}
where $x$ is a variable.
Our aim is to guarantee a maximal uncertainty of approximately $\pm$30\% in the determination of the standard deviation, $\sigma$, depending on the kurtosis value.
The estimated standard deviation of the Gaussian, Laplace and Wigner distributions is computed by generating a data sample, Y(x), for each distribution and then taking the standard deviation of these distributions~:
\begin{align}
    \sigma_Y = \std(Y) =  \ \:  \sqrt{\frac{ \sum_{i=1}^{N} \: (Y_{i} \: - <Y_{i}>)^2  }{N} }\label{eq:express_sigma_for_kurt}
\end{align}
where $\langle .\rangle$ designed an average over the entire data sample, and N is the number of sample.

Figure~\ref{fig_example_kurt_effet_std_val} illustrates that a 30\% uncertainty interval on the determination on $\sigma$ is approximately bounded by $ -1 \leq \kappa \leq 2.7$.
Therefore, to ensure a reliable estimate of the large-scale Alfvén wave fluctuations, we restrict the PSP and SO data computed over $\Tinj$ to those that respect these kurtosis limits. We observe that the mid-high widths of the three distributions differ considerably, while the deviation in their respective sigmas is much smaller. This illustrates that the shape of the entire distribution matters, not just the local width at mid-high, when inferring the statistical spread.

\begin{figure*}[ht]
    \hspace{-0.5cm}
    \includegraphics[width=18.5cm]{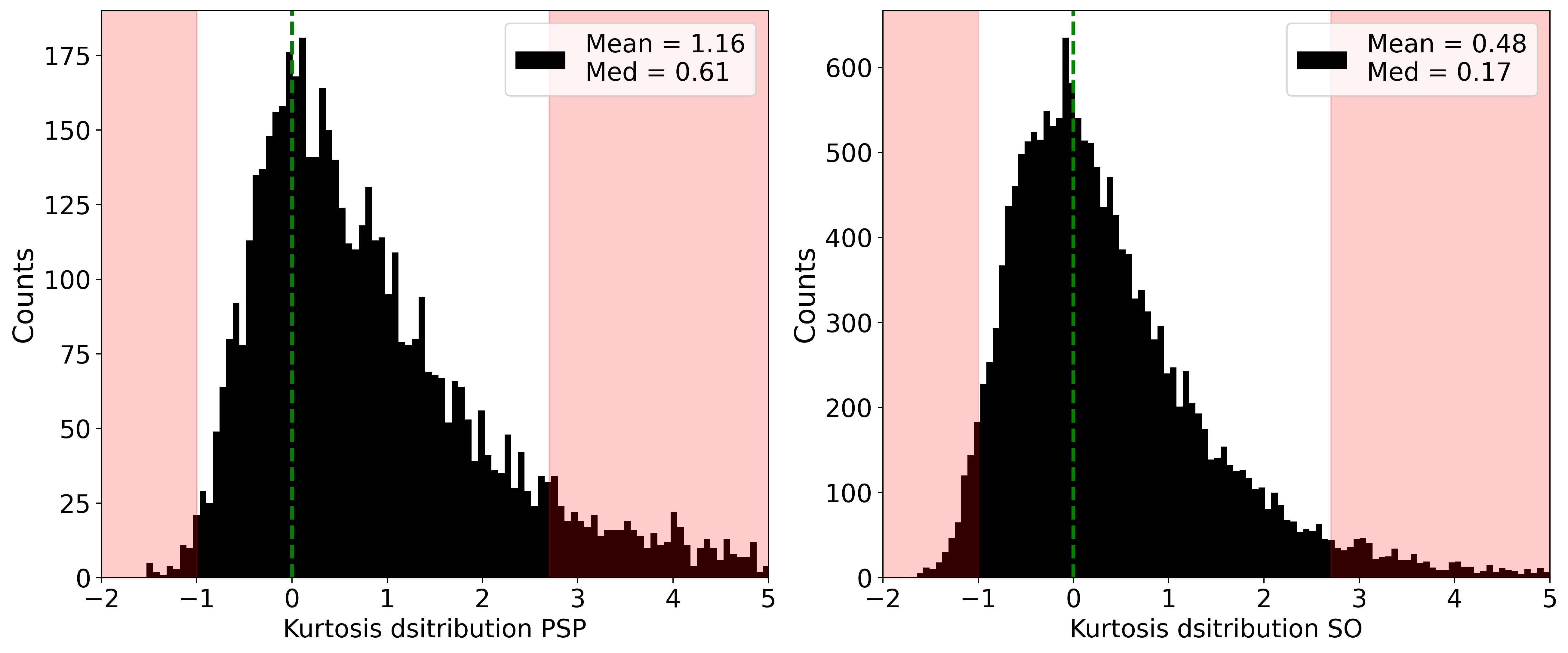}
    \caption{Average excess kurtosis, $\kappa$, computed from Equation~\eqref{eq:express_average_kurtosis} over the time $\Tinj$ for PSP (left) and SO (right) measurements.
    The shaded red regions correspond to excluded data points within each dataset. The green dashed line indicates the excess kurtosis value resulting for an exact Gaussian distribution of EM fluctuations. The median of each kurtosis values distribution are indicated in legends. 
    }
    \label{fig_kurtosis_filt_PSP_SO}
\end{figure*}

Figure~\ref{fig_kurtosis_filt_PSP_SO} shows the excess kurtosis of PSP and SO observations when taken over the observation time $\Tinj$. 
The vast majority of 
excess kurtosis are within the estimated limit of $-1 \leq \kappa \leq 2.7$.
The peak (i.e. the mode) of the kurtosis values distribution, is located near $\kappa = 0$ for both PSP and SO. This indicates that at the chosen timescale $\Tinj$, most of the time EM fluctuation distribution is indeed nearly Gaussian, and that $\sigma$ does well represent fluctuations amplitude at injection scale. 

\subsection{Construction of an Electron Temperature Dataset for Solar Orbiter}
\label{appendix:subsec_electron_SO_creation}

Using observations from PSP, for which electron temperature measurements are available, we compute the relationship between $\Te$ and $v$ using the entire PSP dataset.
Since $\Te$ and $v$ present great anti-correlation \citep[Pearson correlation coefficient of -0.6][]{dakeyo2026}, using $v$ to reconstruct $\Te$ is a coherent approximation. 
A power-law fit is then applied to the ratio $\Te / v$ as a function of heliocentric distance.
The logarithmic correlation between $\Te$ and $v$ yields a Pearson correlation coefficient of $-0.56$ for our PSP dataset, and the ratio follows~:
\begin{equation}
\frac{\Te}{v} = a\, r^{b},
\end{equation}
with $a = 9.83 \pm 0.04$ and $b = -0.79 \pm 0.01$.
Based on these statistical properties, the electron temperature for SO is defined as~:
\begin{equation}
\Te^{(\mathrm{SO})} = \left( a\, r^{b} \right)   v^{(\mathrm{SO})}.
\end{equation}
The resulting synthetic electron temperature for SO is shown in Figure~\ref{fig_u_Tp_Te_n_all_data_med} (d). 
It exhibits good continuity with PSP observations in both radial trend and statistical spread. We therefore conclude that the constructed $\Te$ dataset is suitable for radial trend and energy budget analyses.

\section{Alpha-Particle Influence on the Energy Budget}
\label{appendix:sec:alpha_influence_energy_budget}

Alpha particles are expected to contribute only weakly to the solar wind energy budget. To assess the robustness of this assumption, we estimate their contribution using typical properties: mass $\mal = 4\,m_p$, bulk speed $\val \approx v$, velocity fluctuations %
$\dval \approx \dv$, temperature $\Tal \approx 3.5\times \Tp$, and abundance $\Aal = \nal/n = 0.035$ \citep{Peng2024}.
These approximations aim to emphasize that in average (mixing slow and fast wind), that the main difference between protons and alphas is their respective mass, temperature and density. Even if their respective velocity background and fluctuation are not equal, they are not observed to be different 
as much as other quantities \citep[e.g. $\lesssim 50\%$ of deviation]{Peng2024}. Therefore, velocity properties are estimated to be similar in a lower order approximation.
Including alpha particles modifies the mean particle mass to~:
\begin{equation}
m = (1 - \Aal)m_p + \Aal \mal + m_e \approx 1.09\times m_p.
\end{equation}
This results in an average increase of approximately 9\% in the kinetic and gravitational energies. The thermal energy carried by alpha particles satisfies~:
\begin{equation}
\Ethal \approx 3.5\times \Aal\,\Ethp \approx 0.12\times \Ethp,
\end{equation}
corresponding to an equivalent 12\% increase in the proton thermal energy.
For the wave energy $\Ew$, the contribution varies with total mass %
and therefore increases by approximately 9\%. 
When all energetic contributions are included, the relative increase in $\Etot$ changes from 56\%~($\pm$9\%) to 60\%~($\pm$9\%) with $\Etot \propto r^{0.17 \pm 0.02}$.

\section{Prediction of the average asymptotic wind speed from Ulysses data}
\label{appendix:average_asympt_wind_speed_pred}

Observations of Ulysses \citep{Ulysses_mission_ref}, are used to estimate the average asymptotic speed of the wind speed profile of panel (a) of Figure~\ref{fig_u_Tp_Te_n_all_data_med}. To do so, we use Ulysses proton bulk speed measurements from SWOOP instruments \citep{SWOOP_ref}, taken in between 1998 and 2003 to include a similar solar cycle period than the one covered by PSP and SO (rising up to maximum). Observations are restricted to the same heliospheric latitudes covered by PSP and SO ($|\theta| < 8^\circ $), and separated between $r < 1.5$~au and $r>5$~au, to represent as best as possible the $\sim 1$~au to $5$~au wind speed increase. 
This leads to~: 
\begin{align}
    v_{\:1~\mbox{au}} = \langle v_{r<1.5~\mbox{au}} \rangle 
    = 429 \ \text{km/s}
    \qquad \text{and} \qquad
    \langle v_{r>5~\mbox{au}} \rangle 
    = 468 \ \text{km/s}.
\end{align}

\section{Re-estimation of the gravitational energy flux from Rivera et al. 2024, 2025}
\label{appendix:re-estime_grav_flux_Rivera}

The expression used to compute the gravitational energy flux from \cite{Rivera2024, Rivera2025} is~: 
\begin{align}
    \Wg = \sum_j \:\frac{n_j m_j G M_\odot}{R_\odot} \bigg( 1 - \frac{R_\odot}{r} \bigg)
    \label{eq:appendix_grav_flux_rivera}
\end{align}
where $j$, defines the species (proton, electron or alpha), $n_j$, $m_j$ are respectively the density and mass of the species $j$, and $R_\odot$ is the Sun's radius. While this expressing is mathematically valid, $\Wg$ is the only energy flux within \cite{Rivera2024, Rivera2025} to have a remaining integration constant to be included in it. Indeed, energy flux is computed in between $R_\odot$ up to $r$, thus is decomposes in~:
\begin{align}
    \Wg = - \sum_j \:\frac{n_j m_j G M_\odot}{r} 
    + \underbrace{ \sum_j \:\frac{n_j m_j G M_\odot}{R_\odot} }_{\mathrm{constante}},
    \label{eq:appendix_grav_flux_rivera}
\end{align}
while none of the kinetic ($\Ww$), enthalpy ($\Wh$), Alfvén wave ($\Ww$) or electron heat flux ($\Wqe$) in \cite{Rivera2024, Rivera2025} are expressed including constants of integration. 
To define properly the total energy flux summing all fluxes as
\begin{align}
    \Wtot = \Wk + \Wh + \Wg + \Ww + \Wqe,
    \label{eq:_Wtot_Rivera}
\end{align}
one should either account for all constants of integration on the left hand side of Equation~\eqref{eq:_Wtot_Rivera}, i.e. as being part of $\Wtot$ \citep[e.g. as done in][]{Halekas2023}. Otherwise, the magnitude of $\Wtot$ is not representative of the same flux observed between the two distances $R_\odot$ and $r$. Moreover, gravity provides a vanishing energy contribution at large distances so the effective $\Wg$ is also vanishing.
Hence, we rescale the observed gravitational energy flux from \cite{Rivera2024, Rivera2025} then recalculate $\Wtot$ for the total energy flux to represent exclusively fluxes measured at the distance $r$. This can be done by we multiplying $\Wg$ by the dimensionless quantity $(R_\odot/r)  / (1 - R_\odot/r)$, to obtain the more canonical expression~:
\begin{align}
    \Wg' = - \sum_j \:\frac{n_j m_j G M_\odot}{r} 
    \label{eq:appendix_grav_flux_rivera_corrected}
\end{align}
Using Equation~\eqref{eq:appendix_grav_flux_rivera_corrected}, we defined the corrected total energy flux~: 
\begin{align}
    \Wtot' = \Wk + \Wh + \Wg' + \Ww + \Wqe,
    \label{eq:_Wtot'_Rivera}
\end{align}
Based on the values given in Table S2 of \cite{Rivera2024}, and Tables 3 and 4 of \cite{Rivera2025}, we calculate $\Wg'$, and  $\Wtot'$. 
We summarize the uncorrected (black) and corrected energy flux values (blue) in Table~\ref{tab:re-estimated_Eflux_Rivera}.

\renewcommand{\arraystretch}{1.2}
\begin{table*}[t!]
\centering
\hspace{-1.7cm}
\scalebox{1}{
\begin{tabular}{|l|c|c|c|}
\multicolumn{4}{c}{Fast solar wind stream of \cite{Rivera2024}}
\tabularnewline
    \hline
    Energy Flux Terms × $(r/R_\odot)^2$
    & PSP ($W.m^{-2}$)~: 
    & SO ($W.m^{-2}$)~: 
    & $\Delta (SO - PSP)$
    \tabularnewline
    \hline 
    $\Wk$ & 10.7 $\pm$ 2.0 & 18.8 $\pm$ 1.4 & 8.1 $\pm$ 2.4
    \tabularnewline
    $\Wh$ & 5.70 $\pm$ 0.84 & 0.75 $\pm$ 0.12 & -4.95 $\pm$ 0.85
    \tabularnewline
    $\Wg$ & 25.10 $\pm$ 2.20 & 28.50 $\pm$ 2.10 & 3.40 $\pm$ 3.00
    \tabularnewline
    $\Ww$ & 4.10 $\pm$ 2.90 & 0.25 $\pm$ 0.07 & -3.90 $\pm$ 2.70
    \tabularnewline
    $\Wtot$ & 45.70 $\pm$ 6.60 & 48.00 $\pm$ 3.60 & 2.3 $\pm$ 7.50 \textbf{(+5\% $\pm$16\% )}
    \tabularnewline
    \hline
    \textcolor{blue}{$\Wg'$} 
    & \textcolor{blue}{-2.04 $\pm$ 0.17} 
    & \textcolor{blue}{-0.22 $\pm$ 0.02} 
    & \textcolor{blue}{1.82 $\pm$ 0.19}
    \tabularnewline
    \textcolor{blue}{$\Wtot'$} 
    & \textcolor{blue}{18.46 $\pm$ 3.63} 
    & \textcolor{blue}{19.58 $\pm$ 1.41 } 
    & \textcolor{blue}{1.12 $\pm$ 3.72 \textbf{(+6\% $\pm$ 20\%)}} 
    \tabularnewline
    \hline

    \noalign{\vskip 6pt}
    \multicolumn{4}{c}{Slow solar wind stream of \cite{Rivera2025}}
    \tabularnewline
    \hline
    Energy Flux Terms × $(r/R_\odot)^2$
    & PSP ($W.m^{-2}$)~: 
    & SO ($W.m^{-2}$)~: 
    & $\Delta (SO - PSP)$
    \tabularnewline
    \hline 
    $\Wk$ & 4.76 $\pm$ 0.68 & 11.02 $\pm$ 0.95 & 6.26 $\pm$ 1.17
    \tabularnewline
    $\Wh$ & 3.76 $\pm$ 0.43 & 0.93 $\pm$ 0.08 & -2.83 $\pm$ 0.44
    \tabularnewline
    $\Wg$ & 26.03 $\pm$ 2.08 & 27.87 $\pm$ 2.01 & 1.84 $\pm$ 2.89
    \tabularnewline
    $\Ww$ & 0.73 $\pm$ 0.58 & 0.13 $\pm$ 0.11 & -0.60 $\pm$ 0.59
    \tabularnewline
    $\Wqe$ & 1.54 $\pm$ 0.51 & 1.61 $\pm$ 0.53 & 0.07 $\pm$ 0.74
    \tabularnewline
    $\Wtot$ & 36.82 $\pm$ 3.60  & 41.56 $\pm$ 3.04 & 4.74 $\pm$ 4.30  \textbf{(+13\% $\pm$ 11\%)}
    \tabularnewline
    \hline
    \textcolor{blue}{$\Wg'$} 
    & \textcolor{blue}{-2.10 $\pm$ 0.17} 
    & \textcolor{blue}{-0.22 $\pm$ 0.02} 
    & \textcolor{blue}{1.88 $\pm$ 0.19}
    \tabularnewline
    \textcolor{blue}{$\Wtot'$} 
    & \textcolor{blue}{8.69 $\pm$ 1.13} 
    & \textcolor{blue}{13.47 $\pm$ 1.10 } 
    & \textcolor{blue}{4.78 $\pm$ 1.58  \textbf{(+55\% $\pm$ 18\%)}} 
    \tabularnewline
    \hline

    \noalign{\vskip 6pt}
    \multicolumn{4}{c}{Slow Alfvénic solar wind stream of \cite{Rivera2025}}
    \tabularnewline
    \hline
    Energy Flux Terms × $(r/R_\odot)^2$
    & PSP ($W.m^{-2}$)~: 
    & SO ($W.m^{-2}$)~: 
    & $\Delta (SO - PSP)$
    \tabularnewline
    \hline 
    $\Wk$ & 5.78 $\pm$ 1.53 & 11.54 $\pm$ 2.31 & 5.76 $\pm$ 2.77
    \tabularnewline
    $\Wh$ & 3.10 $\pm$ 0.70 & 0.76 $\pm$ 0.15 & -2.34 $\pm$ 0.72
    \tabularnewline
    $\Wg$ & 19.59 $\pm$ 2.95 & 20.67 $\pm$ 2.80 & 1.08 $\pm$ 4.06
    \tabularnewline
    $\Ww$ & 1.64 $\pm$ 1.10 & 0.73 $\pm$ 0.45 & -0.91 $\pm$ 1.19
    \tabularnewline
    $\Wtot$ & 30.34 $\pm$ 4.24 & 33.70 $\pm$ 2.20 & 3.36 $\pm$ 4.78 \textbf{(+11\% $\pm$ 16\%)}
    \tabularnewline
    \hline 
    \textcolor{blue}{$\Wg'$} 
    & \textcolor{blue}{-1.54 $\pm$ 0.23} 
    & \textcolor{blue}{-0.16 $\pm$ 0.02} 
    & \textcolor{blue}{1.38 $\pm$ 0.25}
    \tabularnewline
    \textcolor{blue}{$\Wtot'$} 
    & \textcolor{blue}{8.98 $\pm$ 2.02} 
    & \textcolor{blue}{12.59 $\pm$ 2.40} 
    & \textcolor{blue}{3.61 $\pm$ 3.10 \textbf{(+40\% $\pm$ 35\%)}} 
    \tabularnewline
    \hline
\end{tabular}
}
\caption{Tables of energy flux budgets from \cite{Rivera2024, Rivera2025}, with gravitational energy flux $\Wg'$ and $\Wtot'$ corrected  
according respectively to Equation~\eqref{eq:appendix_grav_flux_rivera_corrected} and \eqref{eq:_Wtot'_Rivera}. 
\textbf{First table}~: Fast wind stream energy flow budget from Table~S2 of \cite{Rivera2024}, with $r_{PSP} = 13.3~R_\odot$ and $r_{SO} = 127.7~R_\odot$.
\textbf{Second table}~: Slow wind energy flow budget from Table~3 of \cite{Rivera2025}, with $r_{PSP} = 13.35~R_\odot$ and $r_{SO} = 125.0~R_\odot$. 
\textbf{Third table}~: Slow Alfvénic wind energy flow budget from Table~4 of \cite{Rivera2025}, with $r_{PSP} = 13.7~R_\odot$ and $r_{SO} = 130.6~R_\odot$. Original studies values are indicated in black, and modified values in blue. An estimate of the relative variation of $\Wtot$ between PSP and SO is given as percentage in the right column of each table. Uncertainties of $\Wg$ are rescaled by the factor $\Wg'/ \Wg$. Uncertainties of $\Wtot'$ are estimated as the squared sum of all its constituting variables uncertainties. %
According to the original studies results, the electron heat flux $\Wqe$ is only accounted for the slow wind stream of \cite{Rivera2025}.
}
\label{tab:re-estimated_Eflux_Rivera}
\end{table*}

We can see from Table~\ref{tab:re-estimated_Eflux_Rivera} that the values of $\Wg'$ at PSP are of the same order as those presented in the middle panel of Figure 1 from \cite{Halekas2023}.
Regarding $\Wtot'$ values, they respectively increase of +6\% ($\pm$20\%), +66\% ($\pm18\%$) and +40\% ($\pm35\%$) between PSP and SO show, respectively for the fast, slow and slow Alfvénic wind streams studied in \cite{Rivera2024, Rivera2025}. 
The corresponding increase of $\Wtot$ values being +5\%, +13\% and +11\%, respectively, except for the fast wind stream, the slow and slow Alfvénic wind streams show way more than 3 times larger increase when considering the corrected total flux $\Wtot'$. 
Said differently, adding a large constant to $\Wg$ is falsely providing the impression that the total energy is approximately conserved as the percentage of $\Wg$ variation decreases as the constant magnitude is set to a higher value.


\section{Complementary Figures}
\label{appendix:complet_fig}

\begin{figure}[ht]
    \centering
    \includegraphics[width=10cm]{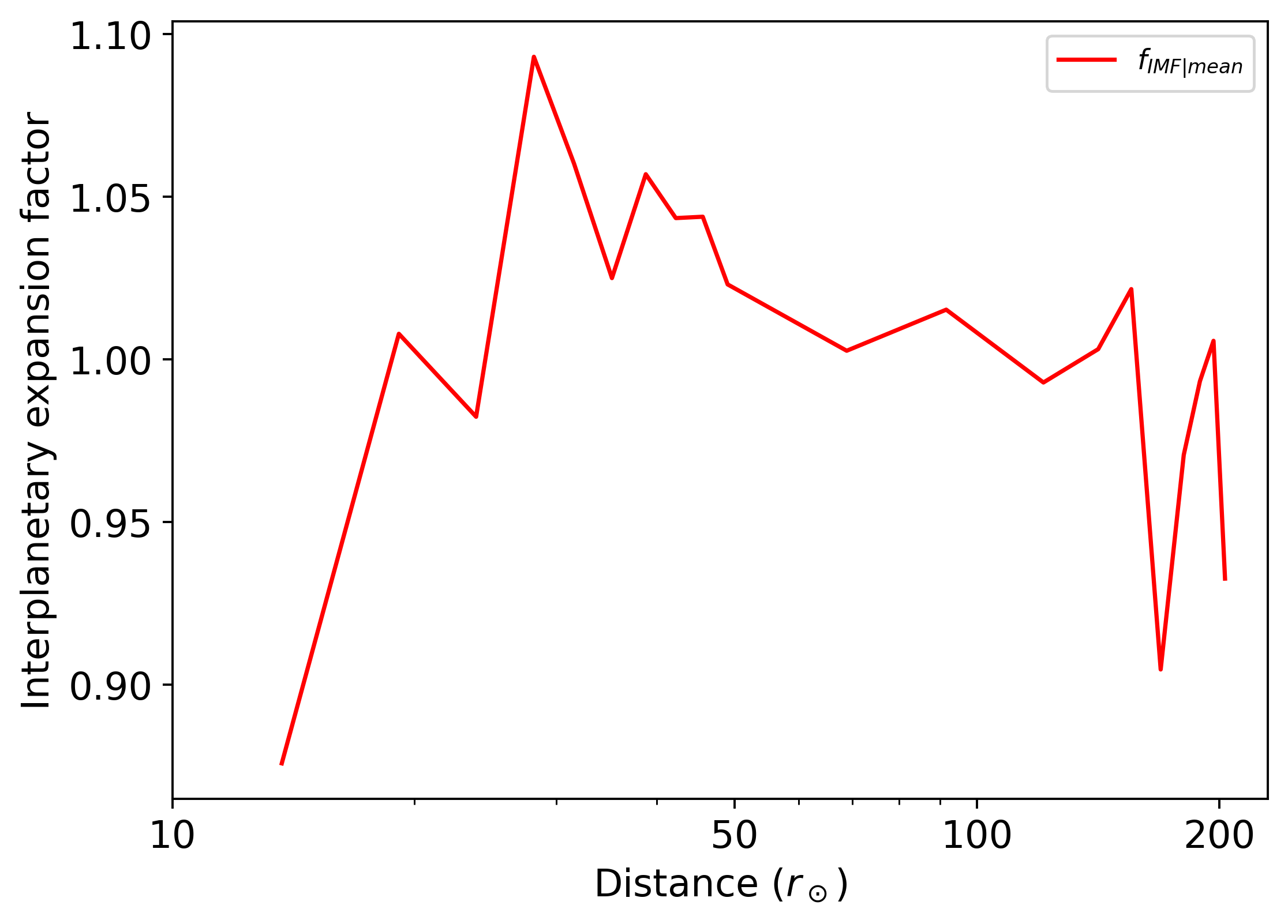}
    \caption{Mean profile of the interplanetary expansion factor, $\fimf$, computed from the deviation to the magnetic flux to spherical expansion with Eq.~\eqref{eq:express_fimf}. 
    The standard error associated to the mean profile (defined in Section~\ref{subsec:description_obs_mean_trend}) is
    $\sigma_{\fimf|err} = 1.6\%$.}
    \label{fig_fimf}
\end{figure}

\begin{figure}[ht]
    \centering
    \includegraphics[width=18.5cm]{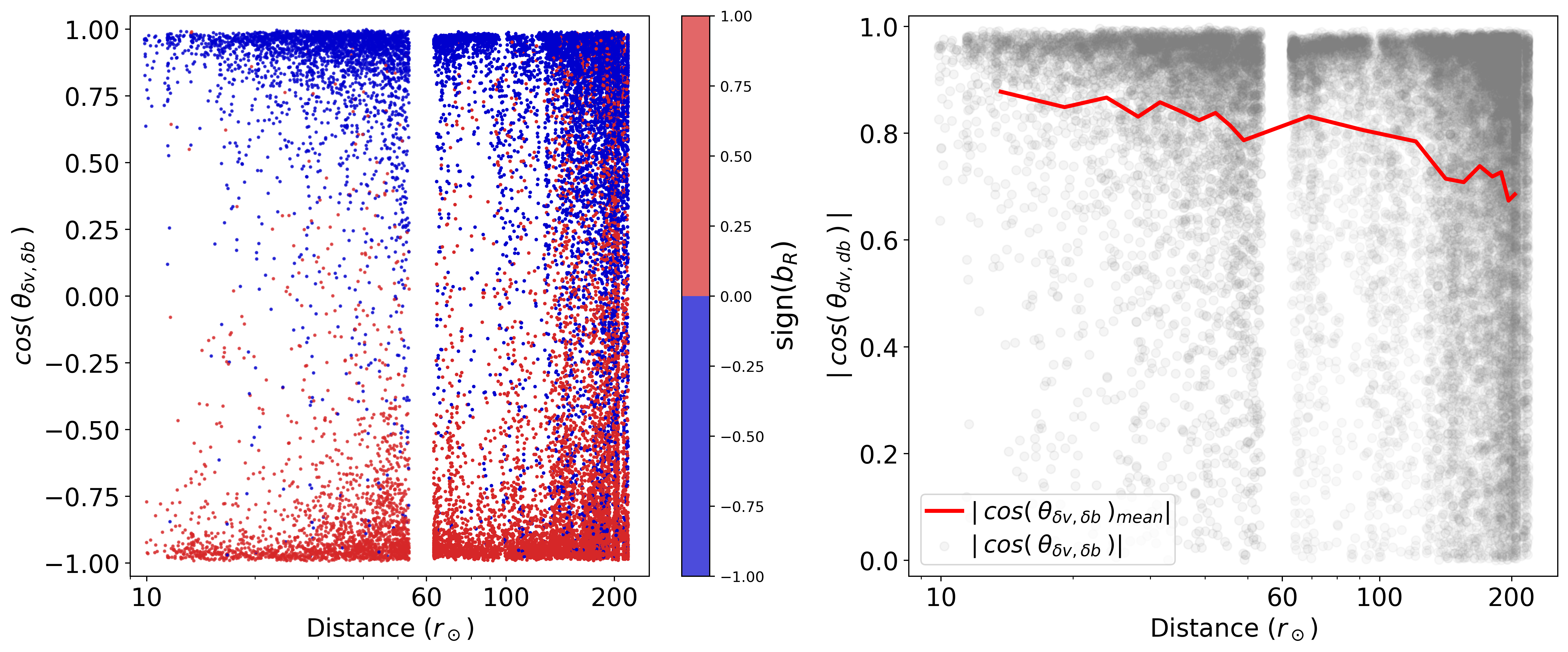}
    \caption{Radial evolution of $\cos(\theta_{\dvvect,\dbvect})$ colored by the radial magnetic field polarity (left), and its absolute value mean profile (right). Accordingly to the definition of a scalar product, the cosine of the angle between $\dvvect$ and $\dbvect$ is defined by $\cos(\theta_{\dvvect,\dbvect}) = \sum_j\dv_j\db_j / \Big(\sqrt{ \sum_j\dv_j^2} \sqrt{ \sum_j\db_j^2} \Big)$, where $j = \{R,N\}$ for PSP, and $j = \{R,T,N\}$ for SO data. The standard error (defined in Section~\ref{subsec:description_obs_mean_trend})  
    associated to the mean profile is $\sigma_{|\cos(\theta_{\dvvect,\dbvect})| \:err} = 1.1\%$.}
    \label{fig_cos_theta_dvdb}
\end{figure}

\end{document}